\documentclass[twocolumn,english,aps,prb,longbibliography]{revtex4-1}
\usepackage{color}
\usepackage{graphicx}
\usepackage[T1]{fontenc}
\usepackage{float}
\usepackage{textcomp}
\usepackage{amsmath}
\usepackage{amssymb}
\usepackage{graphicx}
\usepackage{esint}
\usepackage{natbib}
\usepackage{hyperref}

\begin{document}

\title{Fast evaluation of interaction integrals for confined systems with machine learning}

\author{A. Mre\'nca-Kolasi\'nska}
\affiliation{AGH University of Science and Technology, Faculty of Physics and
Applied Computer Science,\\
 al. Mickiewicza 30, 30-059 Krak\'ow, Poland}

\author{K. Kolasi\'nski}
\affiliation{AGH University of Science and Technology, Faculty of Physics and
Applied Computer Science,\\
 al. Mickiewicza 30, 30-059 Krak\'ow, Poland}

\author{B. Szafran}
\affiliation{AGH University of Science and Technology, Faculty of Physics and
Applied Computer Science,\\
 al. Mickiewicza 30, 30-059 Krak\'ow, Poland}

\begin{abstract}
The calculation of interaction integrals is a bottleneck for the treatment of many-body 
quantum systems due to its high numerical cost. We conduct configuration interaction calculations of the few-electron states confined in III-V semiconductor two-dimensional structures using a shallow neural network to calculate the two-electron integrals, which can be used for general isotropic interaction potentials. This approach allows for a speed-up of the evaluation of the energy levels and controllable accuracy.
\end{abstract}

\maketitle

\section{Introduction}

Computation of quantum problems for interacting particles is a challenging task that has been dealt with using various approaches, including machine learning (ML) \cite{Carleo2017, Carleo2018, Gardas2018, Cai2018, Fabrizio2019, Schutt2019, Grisafi2019, Han2019, Hermann2019, Rigo2020, Liu2017, Nagai2017, Shen2018}. 
The difficulty of accurate many-body calculations arises from the high dimensionality of the space required when taking into account the electron-electron and electron-ion interactions.
Methods developed to solve the many-body quantum problems include the density functional theory (DFT) \cite{Hohenberg1964},  Hartree-Fock method \cite{Hartree1935, Roothaan1960}, and configuration interaction (CI) method \cite{Shavitt1977, Pople1987, Rontani2006, Gimenez2007}. 
These methods contain the computation of the non local potentials due to the particle-particle interaction.
This is a computationally demanding task because of its $\mathcal{O}(n_x^2)$ nature, with $n_x$ being the mesh size. A number of approaches trying to improve the scalability of the non-local potential computation exist, with probably the most common being methods based on the Fourier transform which allow us to reduce the computational complexity to $\mathcal{O}(n_x \log(n_x))$.

The problem of the nonlocal potential calculation has been tackled using plane-wave functions and Gaussian-sum (GS) approximations \cite{Beylkin2009, Genovese2006, Exl2016}.
The authors of Ref. \onlinecite{Genovese2006} developed a method of calculating the kernel via expansion of the density in terms of scaling functions and approximation of the $1/r$ potential in terms of Gaussian functions, which served to avoid the costly three-dimensional integral of the original kernel. 
Reference \onlinecite{Exl2016} proposed an approach to improve the accuracy based on the GS approximation of the kernel with a near-field correction added to account for the discrepancy between the GS-approximated and original kernels. 

On the other hand, calculations for complex quantum systems were done using machine-learning methods. The problem was solved 
through the variational quantum Monte Carlo method \cite{Han2019, Hermann2019} or using ML for effective models \cite{Rigo2020}, which are also used in self-learning Monte Carlo methodologies \cite{Liu2017, Nagai2017, Shen2018}.

In this work we propose an approach to solve the few-electron problem via the CI method. The need to calculate a huge number of Coulomb integrals \cite{Mukherjee1975, Lesiuk2014, Peels2020} is a main bottleneck of this method.
 Although the number of integrals can be reduced by taking into consideration symmetries of the problem and building a basis out of functions that satisfy some constraints (e.g., have the necessary spatial symmetry or spin), the calculation time of Coulomb integrals prevails in this problem, especially in two or more dimensions.
The methods of the calculation of the nonlocal potential developed in Refs.~\cite{Beylkin2009, Genovese2006, Exl2016, Genovese2006, Exl2016} mostly aim to obtain the best precision of the calculations. On the other hand, our objective is to develop an approximate and fast method which can be used to evaluate the energies of few-electron states with a precision sufficient to describe the quantum phenomena in mesoscopic systems.

In this paper we develop a method to calculate the two-electron integrals based on a shallow neural network \cite{Goodfellow-et-al-2016} which can be used for any interaction potential that is isotropic. We present the application of our approach for Coulomb as well as non-Coulomb potentials in one or two dimensions. This method can be extended to three dimensions as well. 
The source code for the implementation of the proposed method is available online \cite{KrzysialkeGithub}.

\section{Methods}

\subsection{Hamiltonian}

We consider $N$ electrons confined in a one- or two-dimensional III-V semiconductor nanostructure.
The Hamiltonian, within the effective-mass approximation, is
\begin{equation}
   H=\sum_{ i=1 }^N h_{i} + \sum_{i=1}^N \sum_{j>i}^N u(\mathbf{r}_i-\mathbf{r}_j), 
\label{eq:dh}
\end{equation}
where $u(\mathbf{r}_i-\mathbf{r}_j)$ is the interaction potential between the $i$th and $j$th electron and $h_i$ is the single-electron Hamiltonian
\begin{equation}
   h_i= -\frac{\hbar^2}{2m^*} \nabla^2_{i} + V(\mathbf{r}_i),
\label{eq:sh}
\end{equation}
with $V(\mathbf{r}_i)$ being the external potential at position $\mathbf{r}_i$. The interaction potential may have various forms, depending on the dimensionality of the system. In thin insulating layers it gets an effective form different from the Coulomb interaction potential due to screening \cite{Cudazzo2011}, and for quasi-one-dimensional systems it has the form of the exponentially scaled complementary error function \cite{1dpot}.
We will focus on isotropic interaction potentials that satisfy $u(\mathbf{r}_i-\mathbf{r}_j)=u(|\mathbf{r}_i-\mathbf{r}_j|)$.

The calculation is performed using the CI method \cite{Mukherjee1975, Rontani2006, Wigner1d} within the basis set formed by Slater determinants $\Phi_k$ constructed from the one-electron wave functions
\begin{eqnarray} 
\Psi_n(\mathbf{r}_1, \dots, \mathbf{r}_N) =& \sum_k c_k^{(n)} \Phi_k(\mathbf{r}_1, \dots, \mathbf{r}_N) 
\\
\label{eq:Slater} \Phi_k(\mathbf{r}_1 , \dots, \mathbf{r}_N ) =& \frac{1}{\sqrt{N!}}\left|
\begin{smallmatrix}
\phi_{k_1}(\mathbf{r}_1 )	&	\dots	& \phi_{k_N}(\mathbf{r}_1 ) \\
\vdots							&	\ddots	&	\vdots					\\
\phi_{k_1}(\mathbf{r}_N )	&	\dots	& \phi_{k_N}(\mathbf{r}_N ) \\
\end{smallmatrix}
\right|
\end{eqnarray}
As the single-electron wave functions $\phi_k$ we use the eigenfunctions of the $h$ operator in the first-order finite-difference approximation on a mesh of points.
The coefficients $c_k^{(n)}$ are found by solving the eigenproblem of the matrix with the elements given by
\begin{equation}
H_{pq} = \langle \Phi_p| H |\Phi_q \rangle.
\end{equation}
Using the Slater-Condon rules, one can show that the matrix elements $H_{pq}$ can be represented in terms of the one-electron and two-electron integrals:
\begin{eqnarray*} 
\begin{aligned}[b]
&\langle i | h | j \rangle = \int d\mathbf{r}_1  \phi^*_i(\mathbf{r}_1 )  h(\mathbf{r}_1 )\phi_j(\mathbf{r}_1 ), \\
&\langle ij|u(\mathbf{r}_1-\mathbf{r}_2 )|kl \rangle \\
&= \int d\mathbf{r}_1  d\mathbf{r}_2 \phi^*_i(\mathbf{r}_1 ) \phi^*_j(\mathbf{r}_2 )  u(\mathbf{r}_1-\mathbf{r}_2 ) \phi_l(\mathbf{r}_2 )  \phi_k(\mathbf{r}_1 ).
\end{aligned}
\end{eqnarray*}
We use a basis formed with Slater determinants built with a finite number of one-electron states $\phi_k$. Their number is determined by verifying, by the convergence of the energies of an $N$-electron system and given a basis containing $n$ states, we have $\left(_N^n \right)$ Slater determinants.

\subsection{Evaluation of the two-electron integrals}
 
The two-electron integral can be written as
\begin{equation}
\langle ij|u(\mathbf{r}_1-\mathbf{r}_2 )|kl \rangle = \int d\mathbf{r}_1  \phi^*_i(\mathbf{r}_1 )  u_{jl}(\mathbf{r}_1 ) \phi_k(\mathbf{r}_1 ),
\end{equation}
where the effective potential $u_{jl}(\mathbf{r}_1 )$ is 
\begin{equation}
\label{eq:eff_pot}   u_{jl}(\mathbf{r}_1 ) = \int d\mathbf{r}_2 \phi^*_j(\mathbf{r}_2 )  u(\mathbf{r}_1-\mathbf{r}_2 ) \phi_l(\mathbf{r}_2 ).
\end{equation}
Denoting the complex function $\rho_{jl}(\mathbf{r}_2) = \phi^*_j(\mathbf{r}_2 )\phi_l(\mathbf{r}_2) $, one can reformulate the integral as a convolution operation
\begin{equation}
\label{eq:conv_inf} u_{jl}(\mathbf{r}_1 ) = \int d\mathbf{r}_2  u(\mathbf{r}_1-\mathbf{r}_2 ) \rho_{jl}(\mathbf{r}_2 ) = (u*\rho_{jl})(\mathbf{r}_1),
\end{equation}
where the asterisk $(*)$ is the convolution operator and $u(\mathbf{r})$ is a nonlocal kernel, e.g., Coulomb interaction $u(r)\approx 1/r$. 

\subsubsection{Evaluating integrals with fast Fourier transform}

The integral in Eq.~(\ref{eq:conv_inf}) can be efficiently evaluated using the fast Fourier transform (FFT) method available in many numerical libraries \cite{mkl}. For example, in the case of a one-dimensional grid of size $n_x$ the computational cost of evaluating a single integral with FFT is 
\begin{equation*}
\mathcal{O}((n_x + P)\log{}(n_x + P)),
\end{equation*}
where $P$ is a padding which is added to both sides of the grid symmetrically. The size of the padding depends on the size of the convolutional kernel. For closed systems (i.e., systems for which $\rho_{jl} = 0$ outside the computational box) with long-range interactions, the size of the kernel is $2n_x+1$, and the size of the padding is equal to the size of the computational box $P=n_x$.
We use this approach as our baseline method. However, in the special case of short-range interactions, the convolution kernel may be truncated, and $P\ll n_x$. In such a case padding has a negligible effect on the computation time. From the above we can see that the evaluation time of the integral in Eq.~(\ref{eq:conv_inf}) can be improved in two ways, (a) truncating the kernel size, which will reduce padding, or (b) using faster implementation of FFT.

In this paper we show that the long-range interaction kernel can be approximated by a series of finite-size kernels for which $P\ll n_x$, so that the total computational time is smaller than the baseline FFT implementation. Additionally, we show that using the existing neural network frameworks \cite{tensorflow2015, pytorch}, we can improve further the computational cost using efficient graphics processing unit (GPU) implementations of the convolution operator. The CPU implementation of the convolution is performed using the fast Fourier transform from the Intel$\textsuperscript \textregistered$ MKL library in the Fortran language and GPU implementation is provided in the Python language and TensorFlow framework.

\subsubsection{Approximating the integral in one dimension}

\begin{figure}
\includegraphics[width=0.8\columnwidth]{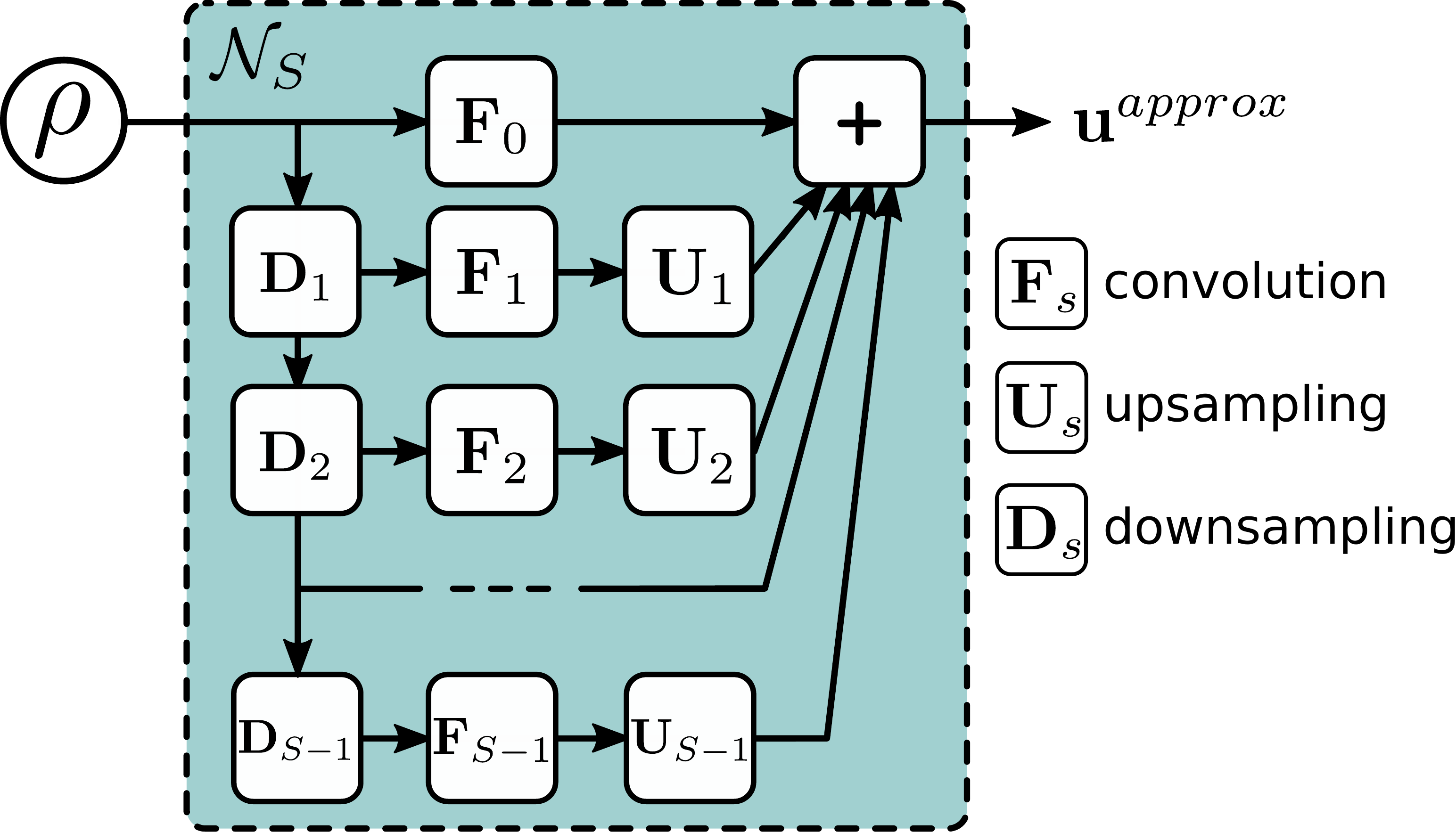}
  \caption{ Schematic of the neural net trained to evaluate the two-electron integrals desribed by Eq.~(\ref{eq:NN}).
  } \label{fig:conv1}
\end{figure}

Let $\boldsymbol{\rho}$ be a discretized one-dimensional (1D) density array of size $n_x$. We define the downsampling operator $\mathbf{D}_s$ as 
\begin{equation*}
\boldsymbol{\hat{\rho}} = \mathbf{D}_s \boldsymbol{\rho},
\end{equation*}
which reduces the spatial size of $\boldsymbol{\rho}$ by a factor of 2 in the $s$th step (for example, starting from $s=0$ with $n_x=512$, operator $\mathbf{D}_1$ downsamples $\boldsymbol{\rho}$ to $n_x=256$, $\mathbf{D}_2$ downsamples $\boldsymbol{\rho}$ from 256 to 128, and so on). In our implementation we use a standard average pool operator, defined as
\begin{equation*}
\hat{\rho}_{i} =\frac{1}{2}(\rho_{2i} + \rho_{2i+1}),\quad i\in(1, n_x/2).
\end{equation*}
Similarly, we define the upsampling operator $\mathbf{U}_s$ as
\begin{equation*}
\boldsymbol{\hat{\rho}} = \mathbf{U}_s \boldsymbol{\rho},
\end{equation*}
which in the $s$th step resizes the input array back to size $n_x$ (for example, for $s=0$ it is an identity operation). In practice, the composition of the downsampling and upsampling operations results in an approximated identity operation
\begin{equation*}
\mathbf{U}_s \mathbf{D}_s \mathbf{D}_{s-1}\dots \mathbf{D}_1 \boldsymbol{\rho} \approx \boldsymbol{\rho}.
\end{equation*}
For the upsampling operation we use the standard bilinear interpolation (e.g. resize\_bilinear operator from the TensorFlow library \cite{tensorflow2015}). The role of the downsampling operation is to increase the receptive field.

We approximate the integral in Eq.~(\ref{eq:conv_inf}) with the following definition of the linear neural network:
\begin{eqnarray}
\label{eq:net_def}
   \mathbf{u}^{approx} = \textbf{F}_0 \boldsymbol{\rho} + \textbf{U}_1\textbf{F}_1\textbf{D}_1 \boldsymbol{\rho} + \textbf{U}_2\textbf{F}_2\textbf{D}_2\textbf{D}_1 \boldsymbol{\rho}  \nonumber \\
\label{eq:NN}   + \dots +  \textbf{U}_{S-1}\textbf{F}_{S-1}\textbf{D}_{S-1}\dots\textbf{D}_2\textbf{D}_1 \boldsymbol{\rho}  \\
   = \mathcal{N}_S(\boldsymbol{\rho}),\nonumber
\end{eqnarray}
where $S$ denotes the number of scales used to approximate the long-range interaction. In each of the $S$ steps, the network downsamples the density, performs a convolution with filter $\mathbf{F}_s$, upsamples the result back to the original mesh size, and sums all the contributions to get the approximate effective potential. Here we assume that $n_x$ is a power of 2 or 
can be divided by 2 at least $S-1$ times, and
 $\textbf{F}_i$ is the convolution operator at the $i$th scale with a learnable kernel of size $K=2 n_x/2^{S-1} + 1$ (we keep the filter size odd). The kernel parameters are not shared between the scales.
 The schematic of the architecture is shown in Fig.~\ref{fig:conv1}. In the next section we describe how to find the optimal parameters $\mathcal{N}_S$.
 
In the special case with the number of scales $S=1$ we have a single convolution with a kernel of size $K=2n_x+1$, and we recover the exact baseline method $\mathbf{u}^{approx} = \mathbf{F}_0 \boldsymbol{\rho} = \mathbf{u}$ described in the previous section. If $S>1$, the computations are no longer exact; however, as we show in Sec.~\ref{sec:appl}, in such a case we can gain a significant improvement in the performance. Note that we do not use any nonlinear activation function in our neural network $\mathcal{N}_S$, Eq.~(\ref{eq:NN}); hence, it preserves the physically required charge superposition condition $\mathcal{N}_S(a\boldsymbol{\rho} + b\boldsymbol{\rho}') = a\mathcal{N}_S(\boldsymbol{\rho}) + b\mathcal{N}_S(\boldsymbol{\rho}')$.

\begin{figure}
\includegraphics[width=\columnwidth]{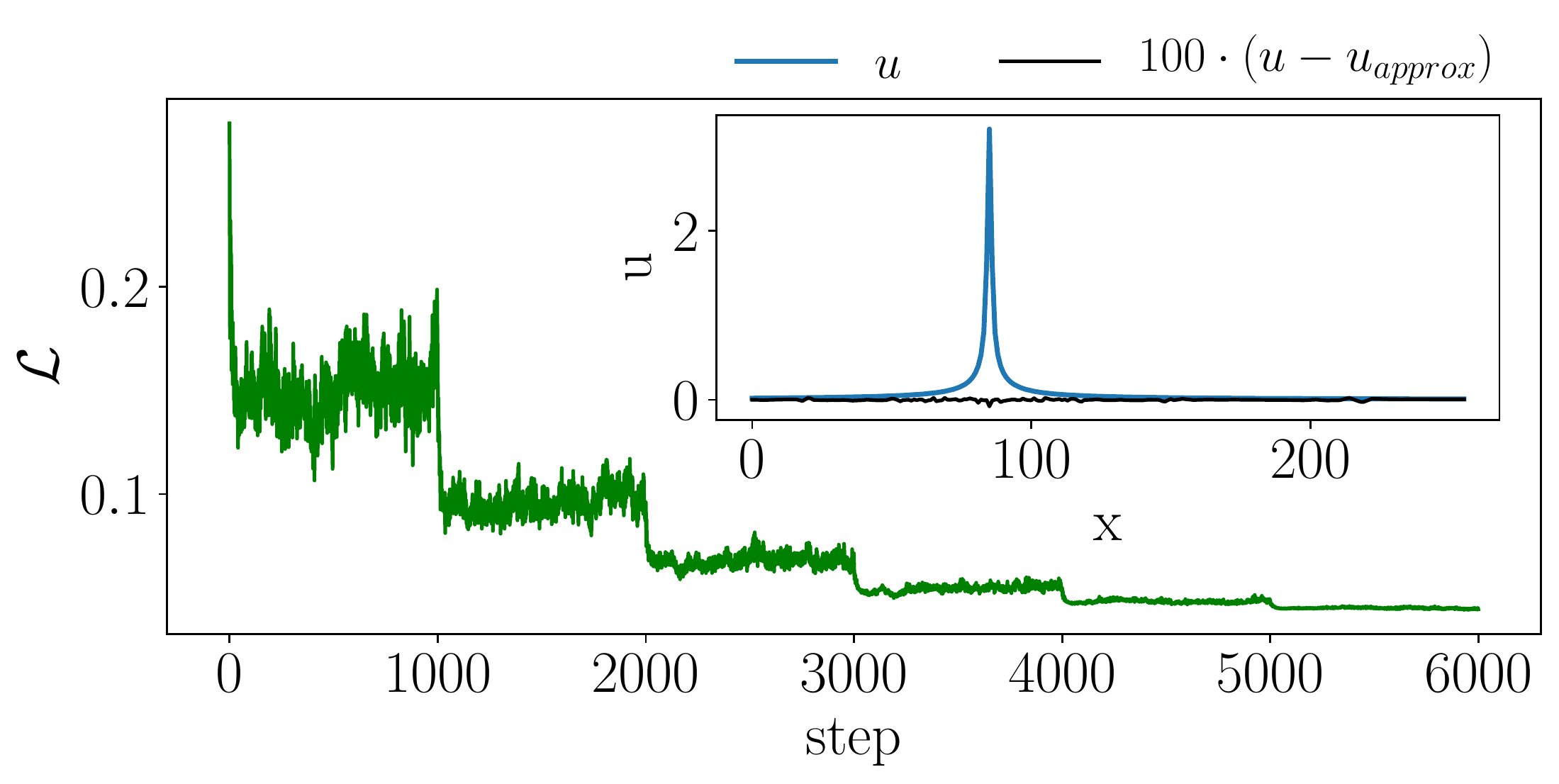}
  \caption{ The loss function across the training iterations. Inset: The potential obtained by the baseline method and the difference between the baseline and approximate result scaled by 100.
  } \label{fig:training}
\end{figure}

\subsubsection{Finding optimal parameters of $\mathcal{N}_S$}
\label{sec:training}

In order to find the optimal kernel parameters of the neural network $\mathcal{N}_S$ we apply the widely used gradient descent (GD) method \cite{Goodfellow-et-al-2016}. We use the TensorFlow library 
\cite{tensorflow2015}, which uses the back-propagation method (i.e., chain rule) to compute the analytical 
value of the gradient of the loss function with respect to the network parameters. In the following we present the loss function and the methodology used to train the network \cite{footnote}.

The effective potential can be treated as a superposition of contributions from point like charges at each mesh point.
We use this property to define the loss function for our problem.
Given a point charge $\rho_p(\textbf{r}) = \delta( \textbf{r}-\textbf{r}_p )$ at position $\textbf{r}_p$, we get the exact solution for the integral (\ref{eq:conv_inf}), and the effective potential is given by the kernel function $u_p (\textbf{r}) = u(\textbf{r}-\textbf{r}_p)$. Substituting Dirac's $\delta$ for the density in Eq.~(\ref{eq:conv_inf}), one obtains
\begin{equation}
\label{eq:delta_dirac}
  u_{p}(\mathbf{r}_1 ) = \int d\mathbf{r}_2  u(\mathbf{r}_1-\mathbf{r}_2 ) \delta(\mathbf{r}_2-\mathbf{r}_p ) = u(\mathbf{r}_1-\mathbf{r}_p).
\end{equation}
To obtain the same result on a discrete grid we use Kronecker's $\delta$ instead of Dirac's $\delta$. Using this property, we train the network $\mathcal{N}_S$ to minimize the difference between the exact $\mathbf{u}_p$ discretized potential and the approximated one [Eq.~(\ref{eq:NN})],

\begin{equation}
\label{eq:train_cost}
 \mathcal{L} = \frac{1}{n_x} \sum_{p=1}^{n_x} \Vert \mathcal{N}_S(\boldsymbol{\rho}_p) - \mathbf{u}_p \Vert^2,
\end{equation}
where $||\mathbf{x}||^2=\sum_i x_i^2$ is the $L_2$ norm of vector \textbf{x} and the sum in the above equation runs over all grid sites. In order to minimize Eq.~(\ref{eq:train_cost}) we use the standard gradient descent using the basic momentum GD optimizer with a decaying learning rate. We decay the learning rate $l_r$ by a factor $\epsilon$ every $N_{it}$ gradient updates. The hyperparameters $l_r$, $\epsilon$, and $N_{it}$ are obtained semi-automatically via a grid search. $N_{it}$ is chosen to be of the order of 2000, and we train the by kernels varying $l_r$ and $\epsilon$ in discrete steps and find their combination that yields the lowest loss function. Note that once the model is trained, it can be reused in many problems assuming that the grid size or the estimated kernel does not change. 

\begin{figure*}
\includegraphics[width=0.8\textwidth]{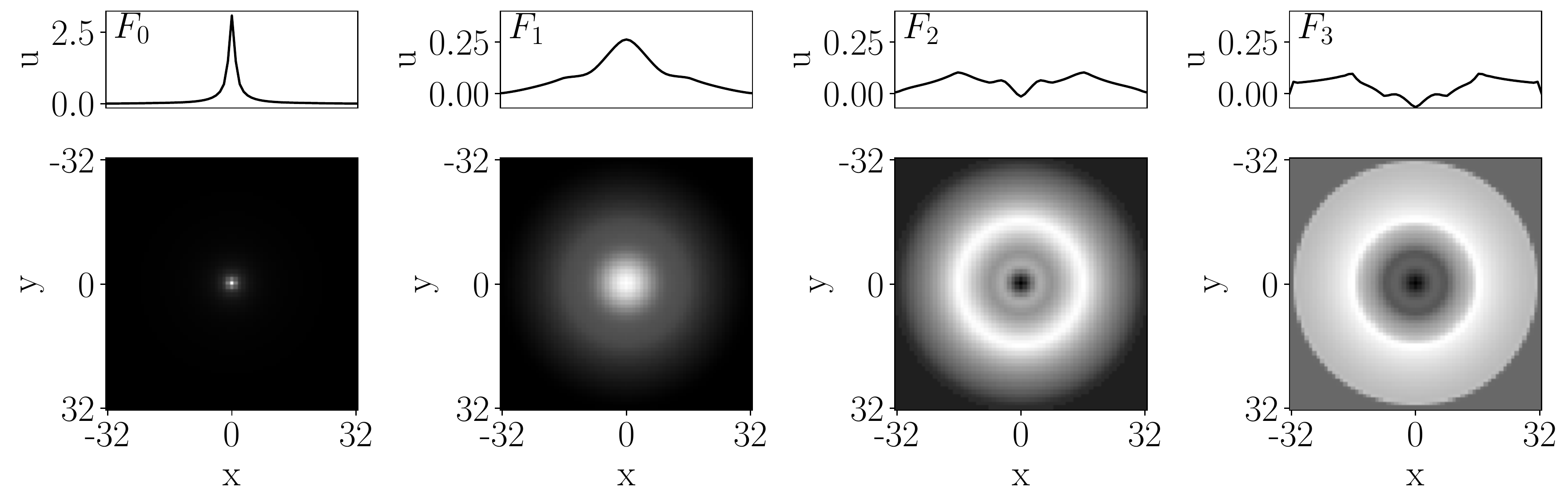}
  \caption{ The kernels obtained from the training in Fig.~\ref{fig:training} using Coulomb interaction potential. Top row: the one-dimensional kernels; bottom row: the kernels after the projection to two dimensions [Eq.~(\ref{eq:projection})].
  } \label{fig:filters}
\end{figure*}

Figure \ref{fig:training} shows the loss function throughout the training for the Coulomb interaction potential on a mesh of $n_x=256$, $S=4$, $K=65$, and initial $l_r=0.003$. The abrupt drop at each multiple of $N_{it}=1000$ steps occurs when the learning rate is decreased by $\epsilon=0.5$.
The potential evaluated by the integral (\ref{eq:eff_pot}) with the baseline method is shown in the inset of Fig.~\ref{fig:training}. The difference between the baseline result and the potential obtained with the trained kernels is shown by the black line. The error is scaled by a factor of 100 to be visible, but the approximated potential is close to the baseline. A more quantitative assessment of the accuracy of our method will follow in Sec. \ref{sec:appl}.

\subsubsection{Isotropic potentials in two dimensions}

In the case of isotropic potentials a one-dimensional kernel $\mathbf{k}^{1D}$ array obtained from the method described in the previous section can be projected to two-dimensional (2D) Cartesian coordinates using the projection tensor $\mathbf{R}$,

\begin{equation}
  k^{2D}_{ij} = \sum_{l=1}^{K} R_{ijl} k^{1D}_{l}, \quad i,j\in(1, K);
  \label{eq:projection}
\end{equation}
for more information about the details of implementation of $\mathbf{R}$ see Ref.~\onlinecite{KrzysialkeGithub}. Having computed the two-dimensional kernels $\mathbf{k}^{2D}$ from Eq.~(\ref{eq:projection}), we can use them to solve the two-dimensional integrals by replacing all the 1D operators in Eq.~(\ref{eq:net_def}) by their 2D analog. A similar approach can be used to project the 1D kernel to three dimensions.

Figure \ref{fig:filters} shows the best kernels at subsequent scales obtained from the training in Fig.~\ref{fig:training}. The 
kernels projected to two dimensions are shown in the bottom row of Fig.~\ref{fig:filters}.

\subsection{Benchmark}

\begin{figure*}
\hspace{0.5cm}(a)\hspace{5.5cm}(b)\hspace{5.5cm}(c)  \\
 \includegraphics[width=0.32\textwidth]{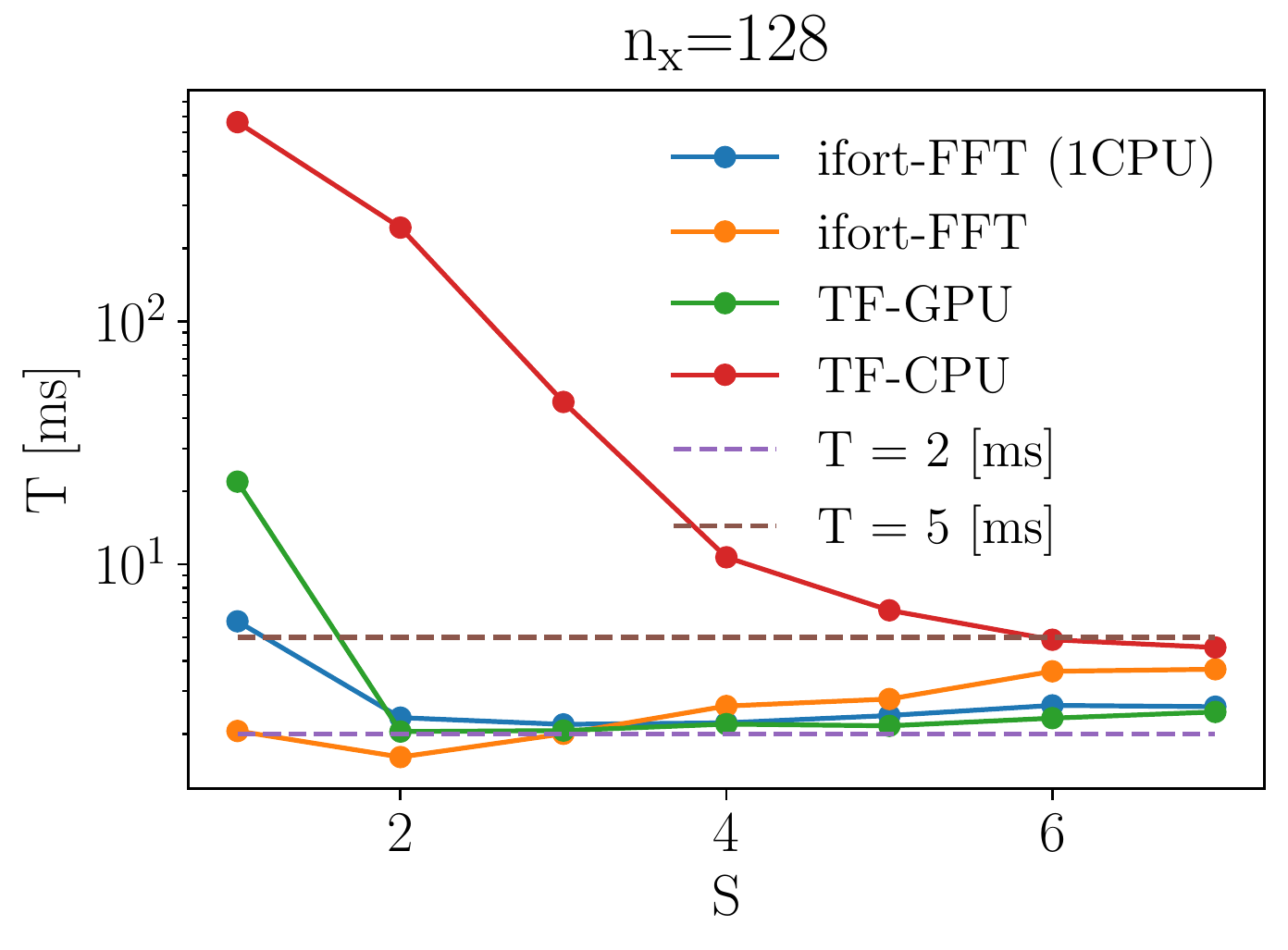}
 \includegraphics[width=0.32\textwidth]{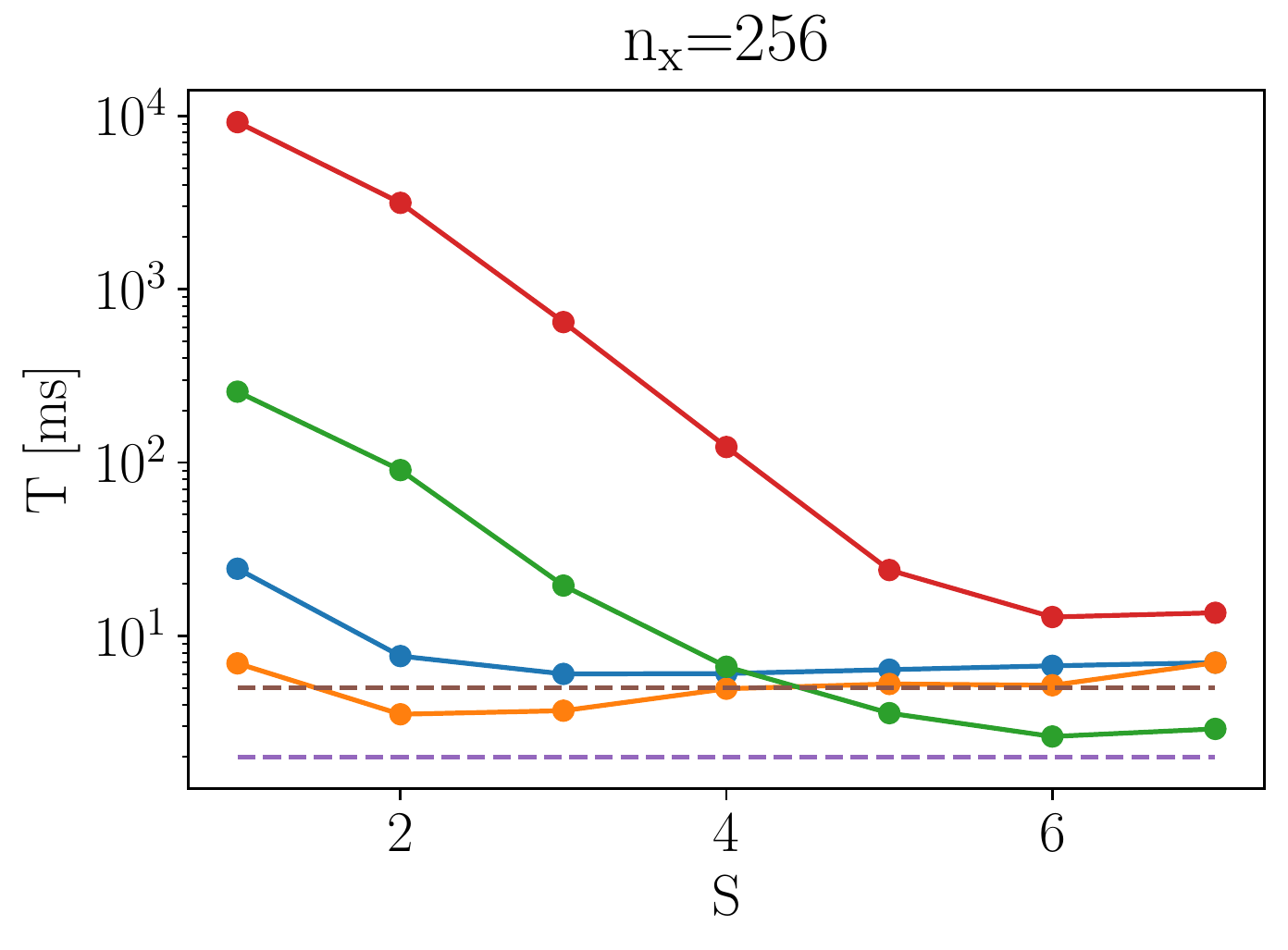}
 \includegraphics[width=0.32\textwidth]{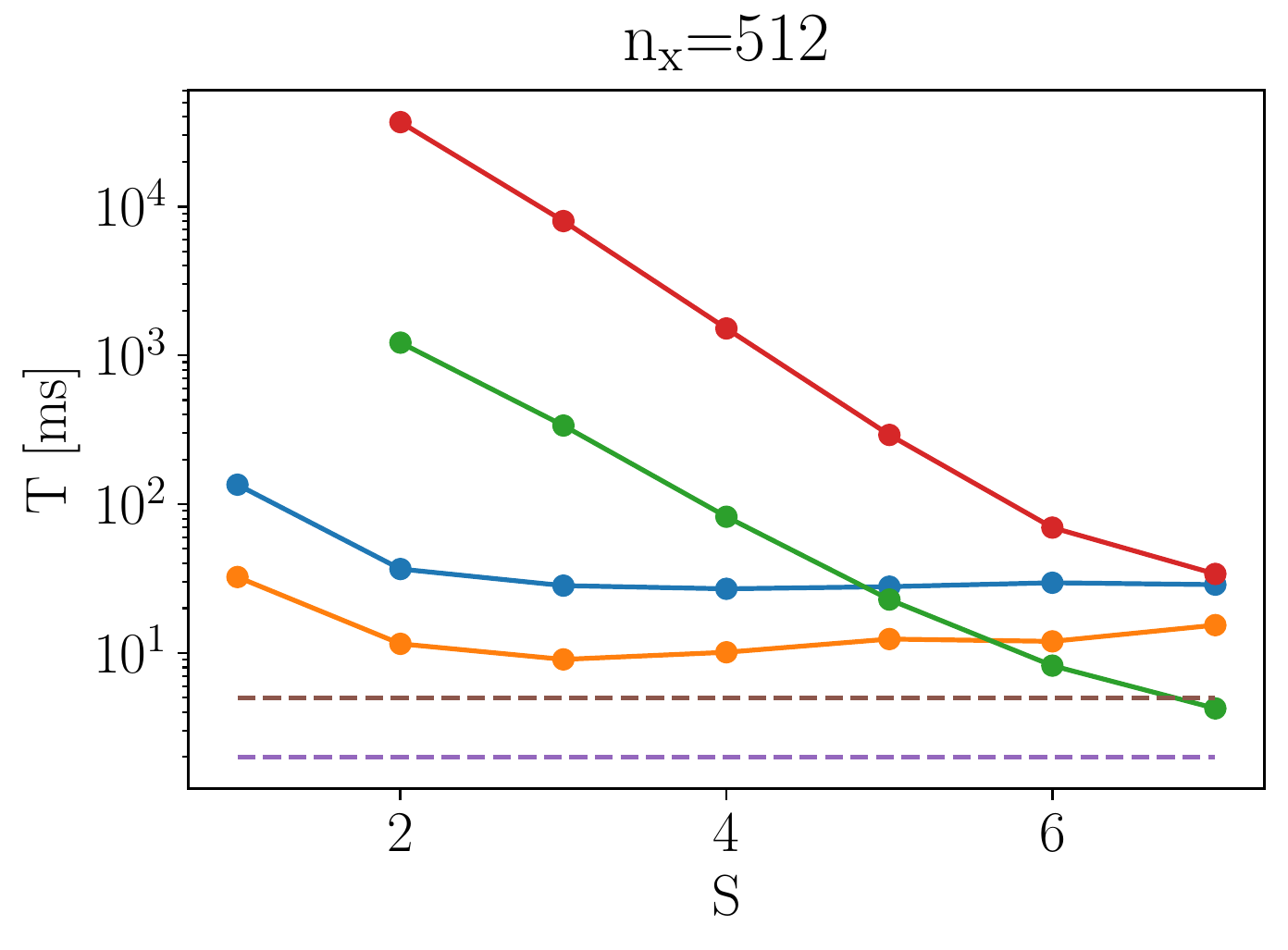}
  \caption{The calculation time averaged over 100 integrals as a function of the number of scales for mesh size (a) $n_x=128$, (b) $n_x=256$ and (c) $n_x=512$.
  } \label{fig:benchmark_n_scales}
\end{figure*}

First, we determine the computation time using the kernels obtained by our method. Figure \ref{fig:benchmark_n_scales} shows the average time of the computation for various mesh sizes and numbers of scales, compared to the baseline. The time includes only the computation of the integrals, and not the training.
We calculate 100 integrals with $\rho$ and kernels filled with random values. The dashed lines show the average time $T=2$ and 5 ms for reference. The computational time is shown for the calculation for a single thread and 40 threads, with the neural net implemented in Fortran, and for the calculation on a CPU and GPU for the TensorFlow implementation. We used the GPU GeForce GTX 1080 Ti. 

The implementation on a GPU works faster for more scales. On the other hand, the MKL implementation is faster with a smaller number of scales, i.e., for the case that is potentially more precise. 

\section{Applications}
\label{sec:appl}
\subsection{Two electrons in a harmonic potential}
As a first application for the method we present the solution of a problem of two interacting electrons confined in a 2D harmonic potential 
\begin{equation}
\label{eq:harmonic} V(r_i) = \tfrac{1}{2} m^* \omega^2 r^2_i,
\end{equation}
with the Coulomb interaction potential 
\begin{equation}
\label{eq:2dVee}
 u(|\mathbf{r}_i-\mathbf{r}_j|) = \frac{e^2}{4\pi\varepsilon_0 \varepsilon } \frac{1}{|\textbf{r}_i-\textbf{r}_j|}.
\end{equation}
We use the GaAs parameters $m^*=0.067m_e$ and $\varepsilon=12.4$, where $m_e$ is the electron mass. We solve the problem using the configuration interaction method with the Coulomb integrals calculated by (i) a convolution with an exact $(2n_x+1)\times (2n_x+1)$ filter by Fourier transform and (ii) using our method.

\begin{figure}
 \includegraphics[width=\columnwidth]{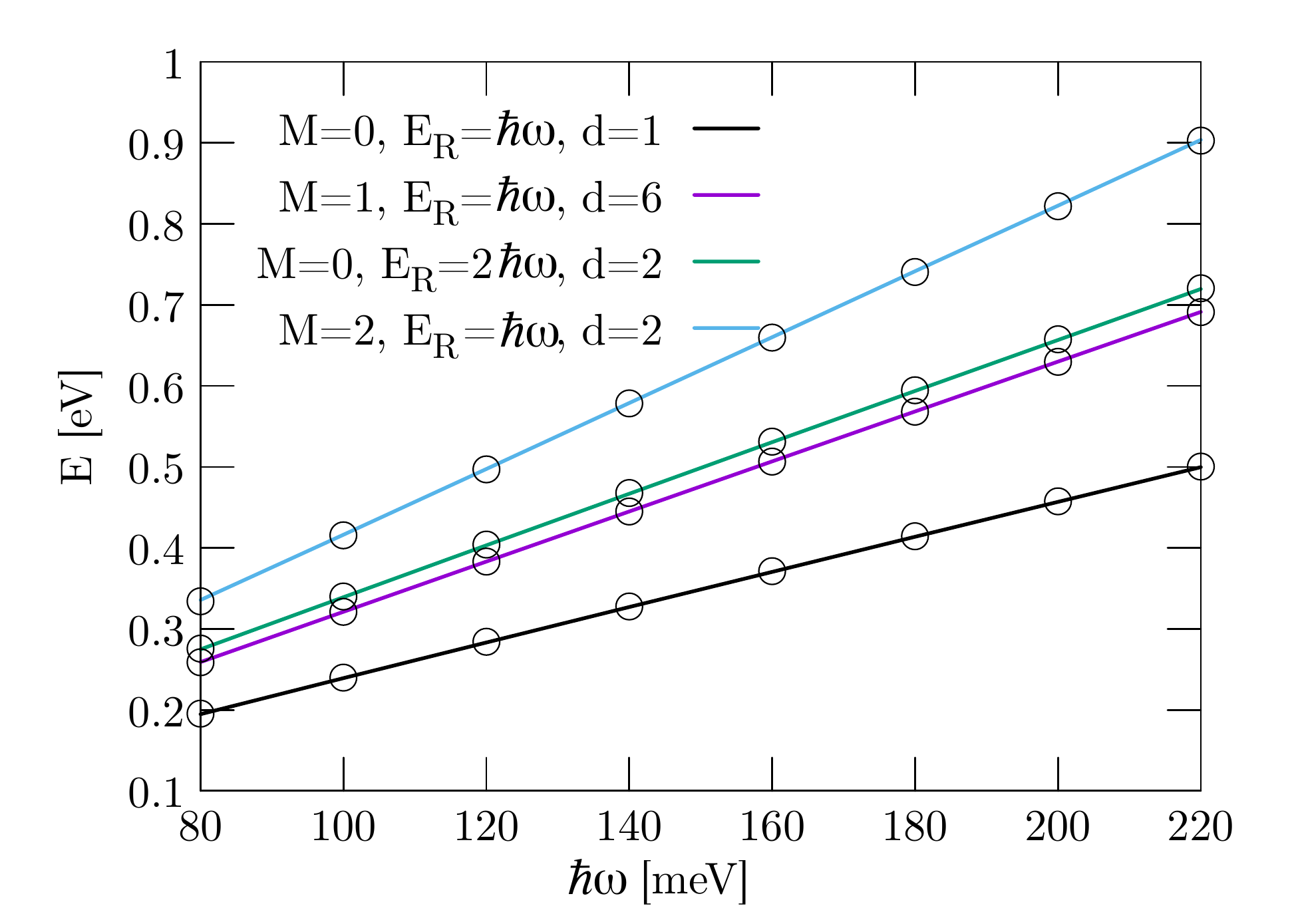}
  \caption{The four lowest energy levels of two electrons trapped in a harmonic potential as a function of $\omega$. The lines show the solution with the shooting method, and the circles show the solution with the CI method with the Coulomb integrals calculated by our method. $M$ is the angular momentum quantum number, and $d$ denotes the degeneracy of the levels.
  } \label{fig:omega_harmonic}
\end{figure}

This problem can also be solved for two electrons using the semianalytical method described in Ref. \onlinecite{SZAFRAN1999}. In the center-of-mass coordinates the Hamiltonian can be written in the form 
\begin{equation}
   H=H_{cm}+H_{rel},
\label{eq:dh_CM}
\end{equation}
where $H_{cm}$ is the center-of-mass Hamiltonian and $H_{ref}$ describes the relative motion of the electrons. $H_{cm}$ is independent of the interaction, and the center-of-mass energy is $E_{R}=\hbar \omega(n_R^x+n_R^y+1)$, where $n_R^x, n_R^y$ are the quantum numbers of the center-of-mass energy.
Further noting that $H_{ref}$ commutes with the $z$ component of the angular momentum operator, one can write it in the cylindrical coordinates
\begin{equation}
   H_{rel,\rho} = -2 \left( \frac{d^2}{d\rho^2} +\frac{1}{\rho}\frac{d}{d\rho} - \frac{M^2}{\rho^2} - \frac{\gamma^2}{4}\rho^2-\frac{1}{\rho} \right),
\label{eq:dh_CM_polar}
\end{equation}
with $\gamma=\omega/2$, which yields states with a well defined angular momentum quantum number $M$. Equation (\ref{eq:dh_CM_polar}) is written in donor units with energy in $R_D=\frac{m^*\kappa^2 e^4}{2\hbar^2\varepsilon^2}$, length in units of $a_D=\frac{\hbar^2\varepsilon}{m^*\kappa e^2}$, and $\kappa=1/4\pi\varepsilon_0$. 
Further, $E=E_{rel}+E_R$. We calculate $E_{rel}$ using the shooting method (see Appendix \ref{shooting}). The four lowest levels and their degeneracies $d$ are given in Fig.~\ref{fig:omega_harmonic}.

\begin{figure}
  \includegraphics[width=\columnwidth]{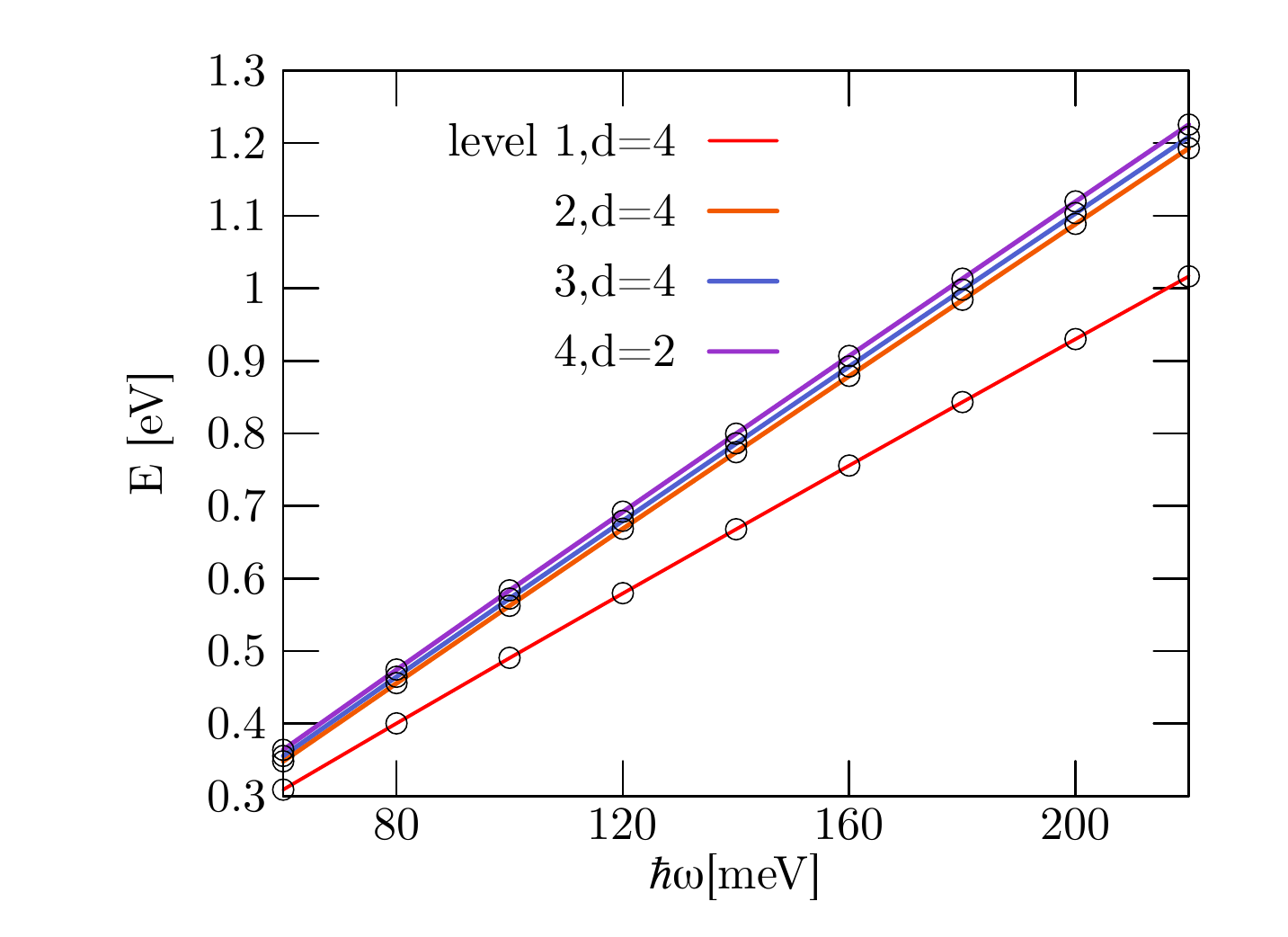}
  \caption{The four lowest energy levels of three electrons trapped in a harmonic potential as a function of $\omega$. The lines show the numerical solution with method (i), and the circles show that with method (ii). $d$ denotes the degeneracy of the levels.
  } \label{fig:omega_harmonic3}
\end{figure}

For the CI method we take as a basis set $n_{basis}$=20 spin orbitals. 
We solve the problem on an $n_x\times n_x$ mesh. 
The results of the calculation with $n_x=64$ are shown in Fig.~\ref{fig:omega_harmonic} together with the results of the shooting method. The results of both methods agree very well. For completeness, we show the results of methods (i) and (ii) for three electrons in Fig.~\ref{fig:omega_harmonic3}.

We present the performance of our method as $n_x$ is varied from 64 to 512, doubling $n_x$ at each step. The results are obtained for $\hbar \omega=200$ meV.
 The hyperparameters used for each mesh size are summarized in Table \ref{tab:hyperparams}.
 
\begin{table}[htbp]
\begin{center}
\begin{tabular}{|c|c|c|c|c|c|c|}
\hline
\multicolumn{7}{|c|}{Interaction potential Eq.~(\ref{eq:2dVee}), 2 dimensions} \\
\hline
   $n_x$	& $S$		& $K$	& $l_r$		& $\epsilon$	&	$N_{it}$	& 	$\mathcal{L}$	\\
\hline
   64	&	2		& 65	& 0.004		& 	0.2			&	2000		& 	0.0122 \\
   128	&	3		& 65	& 0.007		& 	0.4			&	2000		& 	0.0212 \\ 
   256	&	4		& 65	& 0.003		& 	0.4			&	2000		& 	0.0271 \\
   512	&	5		& 65	& 0.002		& 	0.4		&	2000		& 	0.0293	\\
\hline
\end{tabular}
\end{center}
\caption{Hyperparameters used for training the kernels for Fig.~\ref{fig:time_harmonic} for each mesh of size $n_x\times n_x$ and the obtained loss function $\mathcal{L}$ for the interaction potential (\ref{eq:2dVee}) in two dimensions.}
\label{tab:hyperparams}
\end{table}

\begin{figure}
\hspace{0.5cm}(a)\hspace{4cm}(b) 
 \includegraphics[width=0.49\columnwidth]{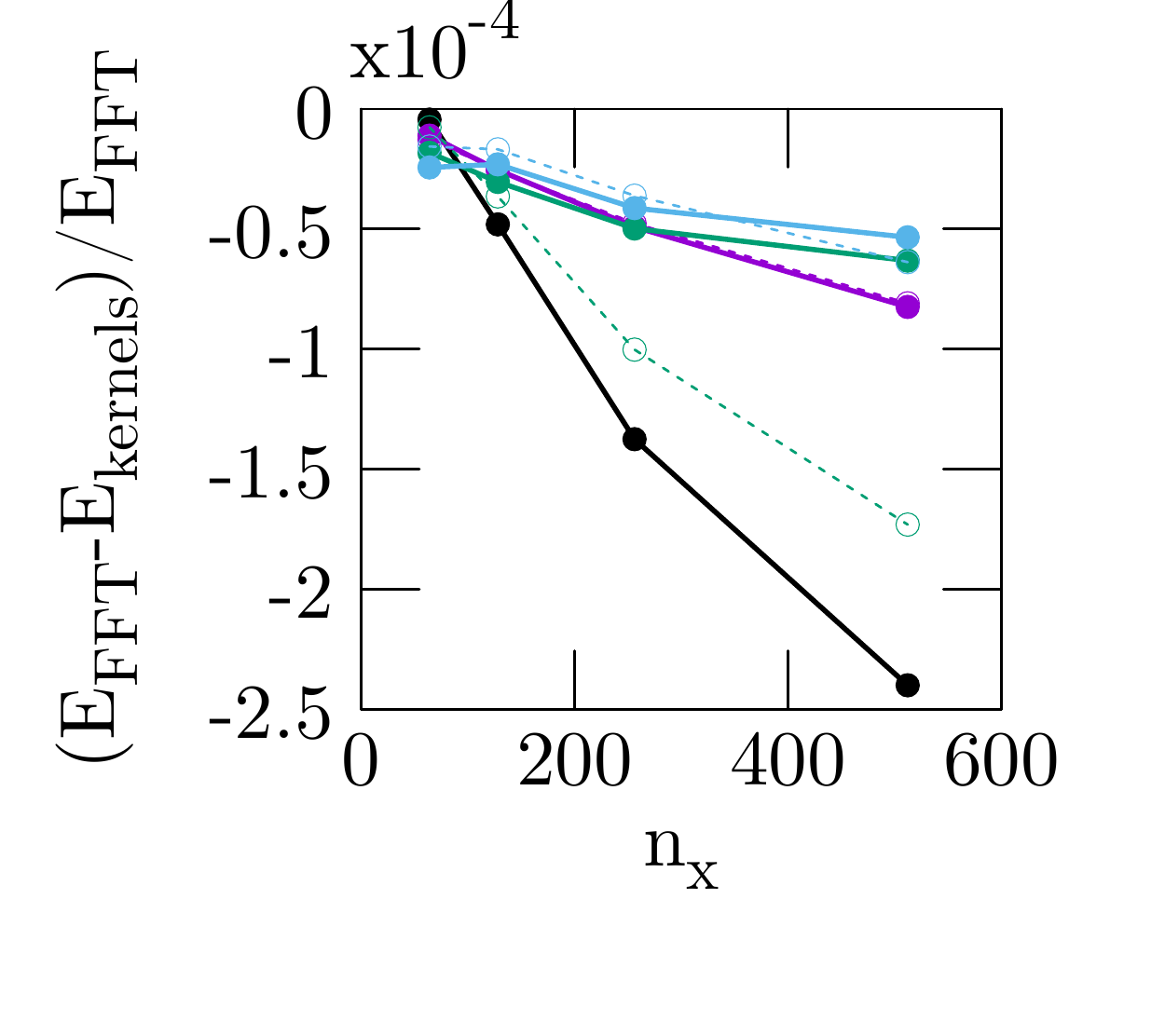}
 \includegraphics[width=0.49\columnwidth]{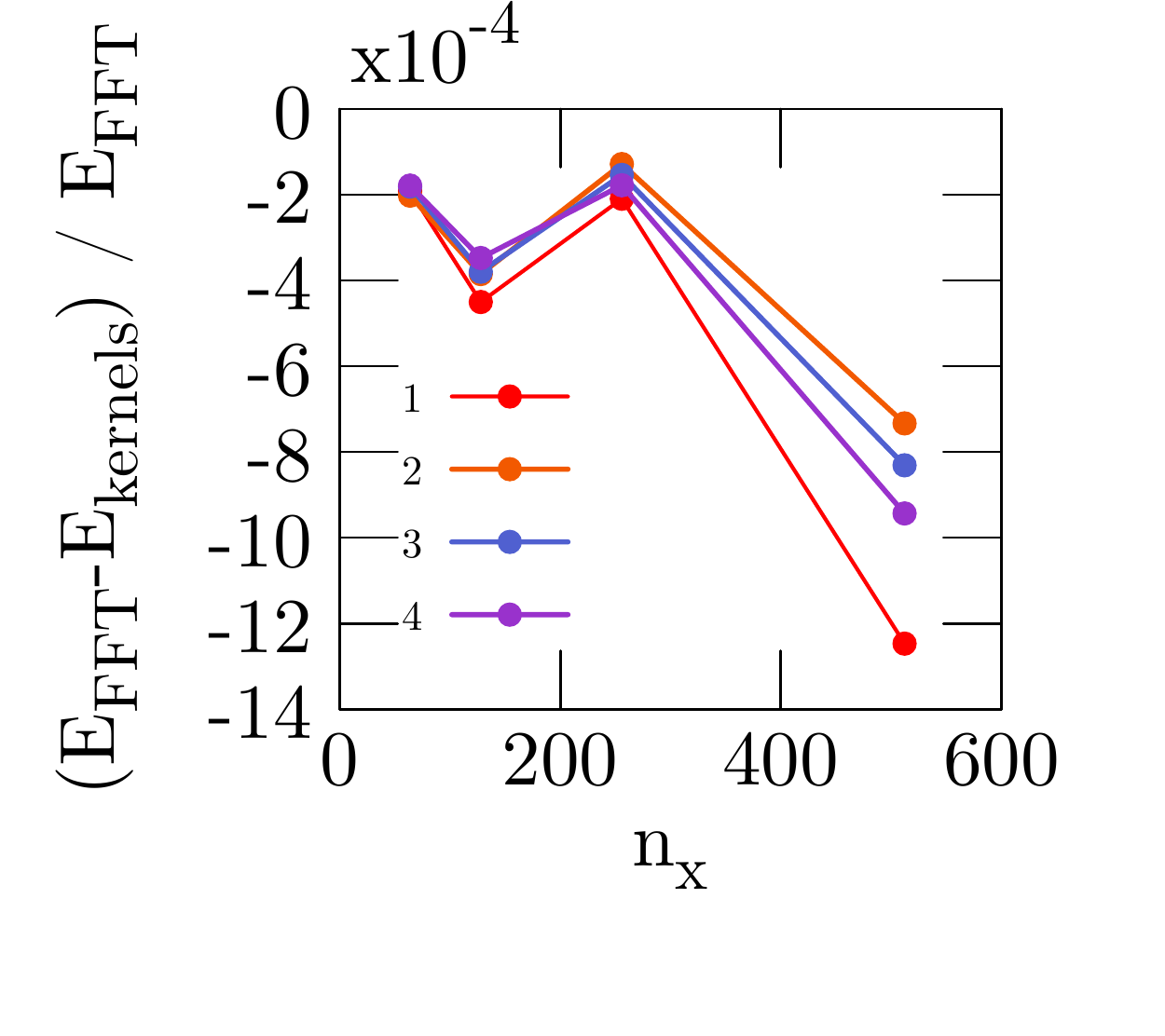}
\caption{The difference between energies obtained with both methods relative to the result of method (i) for the respective energy levels shown in the same line color as in Figs.~\ref{fig:omega_harmonic} and \ref{fig:omega_harmonic3}, for (a) two electrons and (b) three electrons.
  } \label{fig:time_harmonic}
\end{figure}

Figures \ref{fig:time_harmonic}(a) and (b) show the difference between the energies obtained with both methods relative to the result of method (i). Method (i), although not exact, will be more accurate than the approximation of the integral with the sum of scaled convolutions, so we treat it as a reference. Our approach 
gives energies that are relatively close to the baseline result, and the difference is of the order of $10^{-4}$ of the baseline energy.
For energies on the scale of hundreds of meV (see Fig.~\ref{fig:omega_harmonic}) the difference is impossible to spot. 

\begin{figure}
\hspace{0.5cm}(a)\hspace{4cm}(b) 
 \includegraphics[width=0.49\columnwidth]{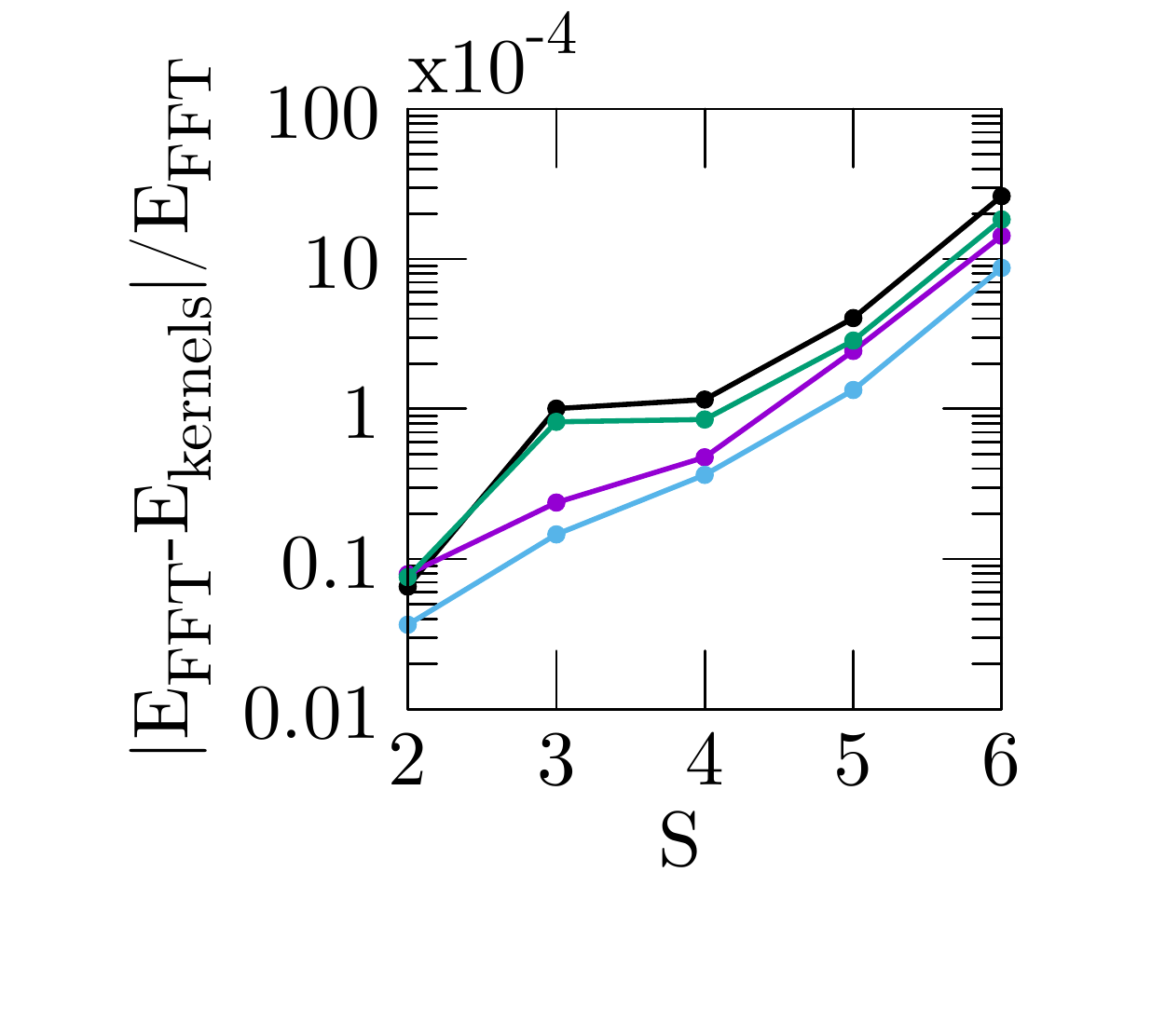}
  \includegraphics[width=0.49\columnwidth]{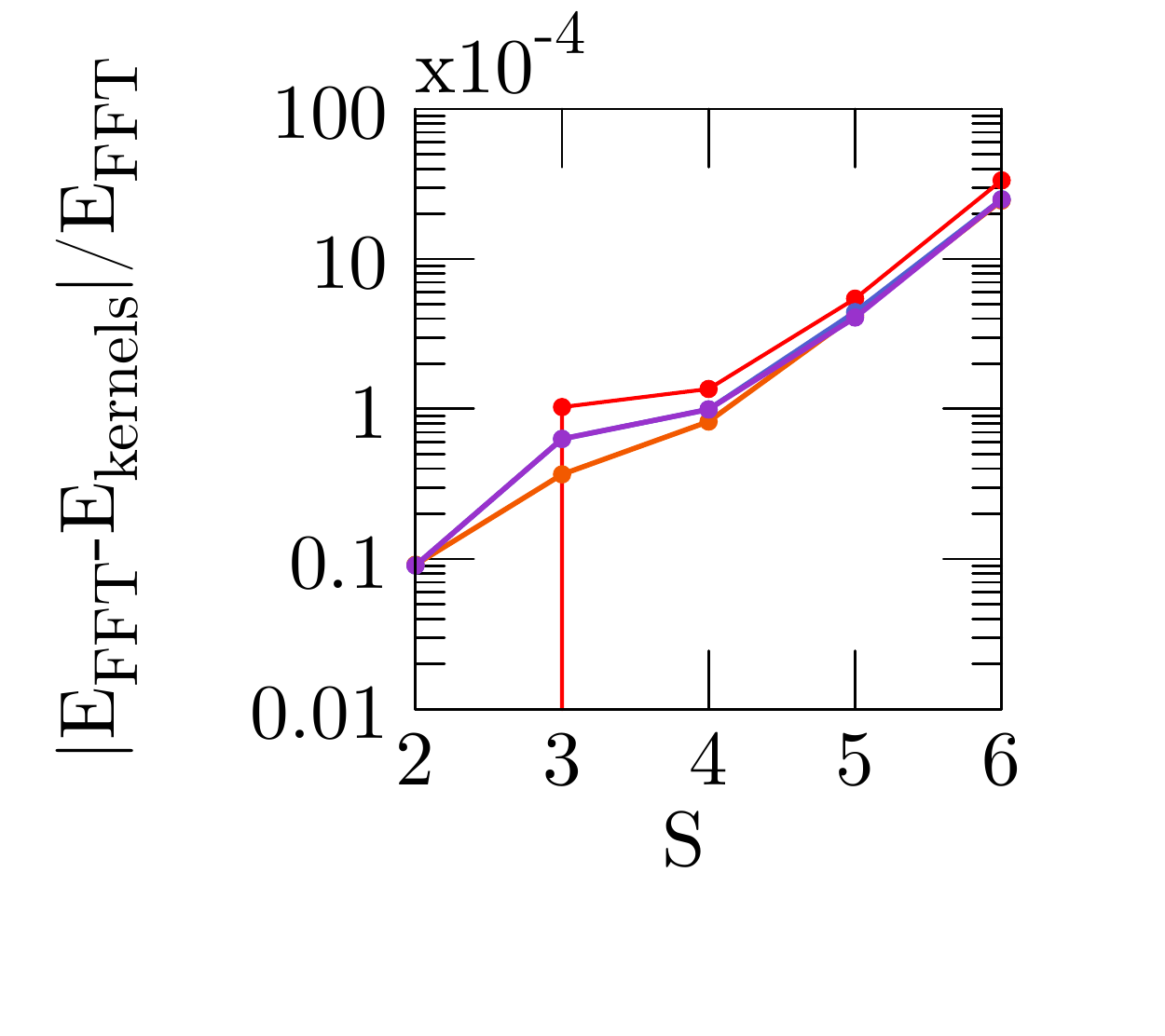}
  \caption{The difference between energies obtained with both methods relative to the result of method (i) for mesh size $n_x=256$ as a function of the number of scales.
 The results are shown for the respective energy levels with the same line color as in Fig.~\ref{fig:omega_harmonic}. The results are obtained for (a) two electrons and (b) three electrons. 
The lines are guides to the eye.
  } \label{fig:error_scales256}
\end{figure}

We consider the accuracy of the method depending on the number of scales $S$ and size of the kernel $K$. Figure \ref{fig:error_scales256} shows the relative error as a function of the number of scales for a $256\times 256$ mesh. For $S=$2, 3, 4, 5, and 6, the filter sizes are $K=$257, 129, 65, 33, and 17, respectively. The parameters used for training the kernels are summarized in Table \ref{tab:hyperparams2}.

\begin{table}[htbp]
\begin{center}
\begin{tabular}{|c|c|c|c|c|c|c|}
\hline
\multicolumn{7}{|c|}{Interaction potential Eq.~(\ref{eq:2dVee}), 2 dimensions} \\
\hline
   $n_x$	& $S$		& $K$	& $l_r$		& $\epsilon$	&	$N_{it}$	& 	$\mathcal{L}$	\\
\hline
   256	&	2		& 257	& 0.008		& 	0.2			&	2000		& 	0.0050 \\
   256	&	3		& 129	& 0.010		& 	0.2			&	2000		& 	0.0142 \\ 
   256	&	4		& 65	& 0.005		& 	0.3			&	2000		& 	0.0267 \\
   256	&	5		& 33	& 0.008		& 	0.3			&	2000		& 	0.0549	\\
   256	&	6		& 17	& 0.004		& 	0.3			&	1800		& 	0.1251	\\
\hline
\end{tabular}
\end{center}
\caption{Hyperparameters used for training the kernels for Fig.~\ref{fig:error_scales256} for a mesh of $256\times 256$ for each $K$ and the obtained loss function $\mathcal{L}$ for the interaction potential (\ref{eq:2dVee}) in two dimensions.}
\label{tab:hyperparams2}
\end{table}

As can be expected, for smaller kernels (and a higher number of scales $S$) the error increases. However, even for the smallest kernel size, the errors do not exceed $4\times 10^{-3}$ of the reference energies. The benchmark in Fig.~\ref{fig:benchmark_n_scales} shows that the method tends to be faster for a larger $S$. Thus choosing the hyperparameters $K$ and $S$ is a trade-off between speedup and accuracy.

\subsection{Effective 1D interaction potential}
 
As the next example of the application of our approach
we present the results for the integration of a non-Coulomb interaction potential in 1D systems. 
We consider a quasi-one-dimensional quantum dot, formed in a semiconductor by strong confinement in two directions \cite{Wigner1d}.
We assume a harmonic oscillator confining potential in the ($x$, $y$) direction. Assuming that for a strong lateral confinement the electrons are frozen to the ground harmonic potential state and integrating over the lateral coordinates, one obtains the interaction potential \cite{1dpot}
 
\begin{equation}
u(z_{ij}) = \left( \pi/2 \right)^\frac{1}{2} (\kappa/l)  \text{erfc}(z_{ij}/2^\frac{1}{2}l )\exp(z_{ij}^2/2l^2).
\label{eq:1dVee}
\end{equation}
Here $z_{ij}=|z_i-z_j|$, and $l=\sqrt{\hbar/m^*\omega}$. 
The single-electron Hamiltonian (\ref{eq:sh}) is 

\begin{equation}
 h_i= -\frac{\hbar^2}{2m^*} \frac{d^2}{dz_i^2} + V(z_i),
\label{eq:1d_sh}
\end{equation}
and we assume a 1D infinite well confinement potential in $V(z_i)$.

\begin{figure}
\hspace{0.5cm}(a)\hspace{4cm}(b) \includegraphics[width=0.49\columnwidth]{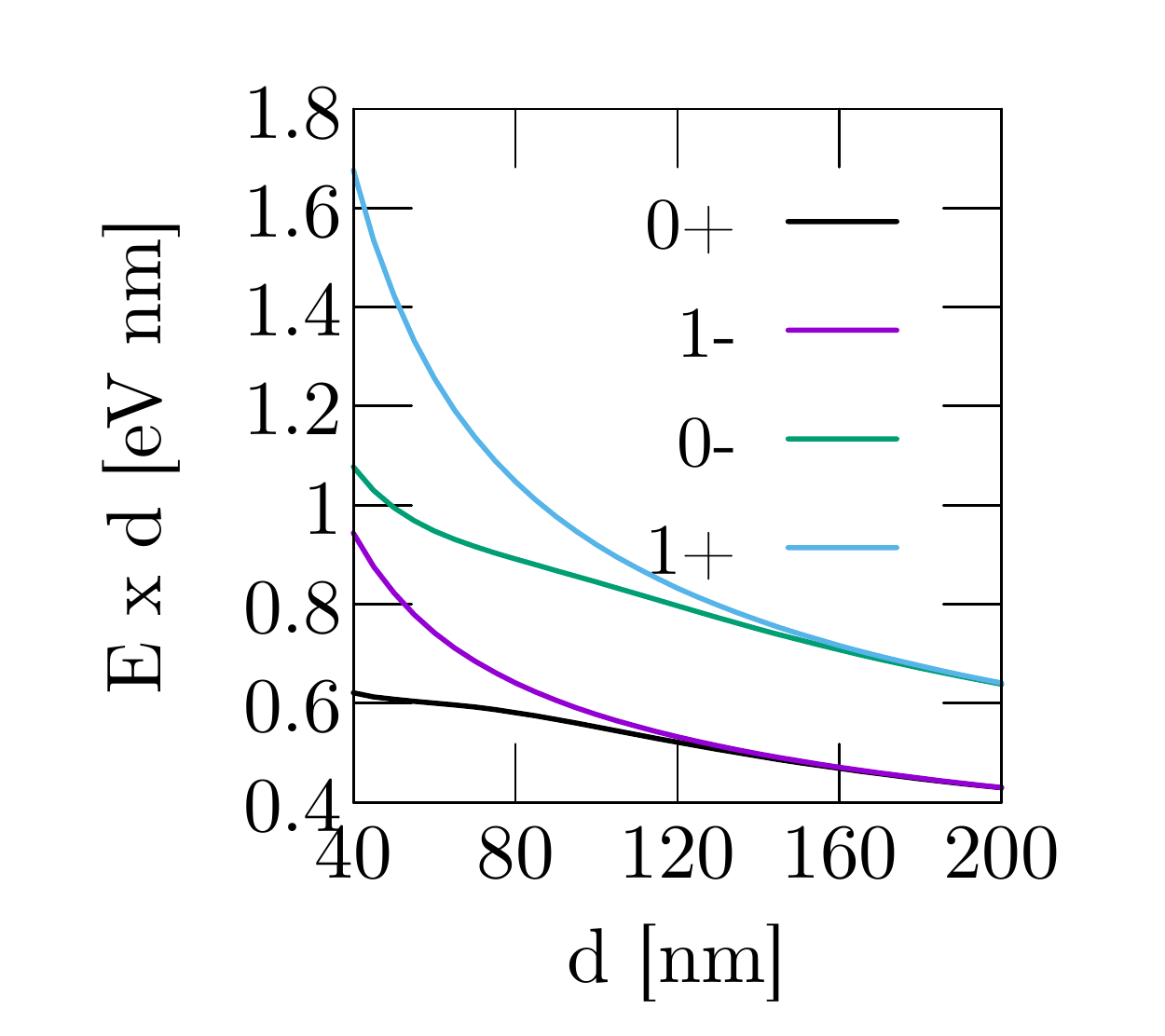}
\includegraphics[width=0.49\columnwidth]{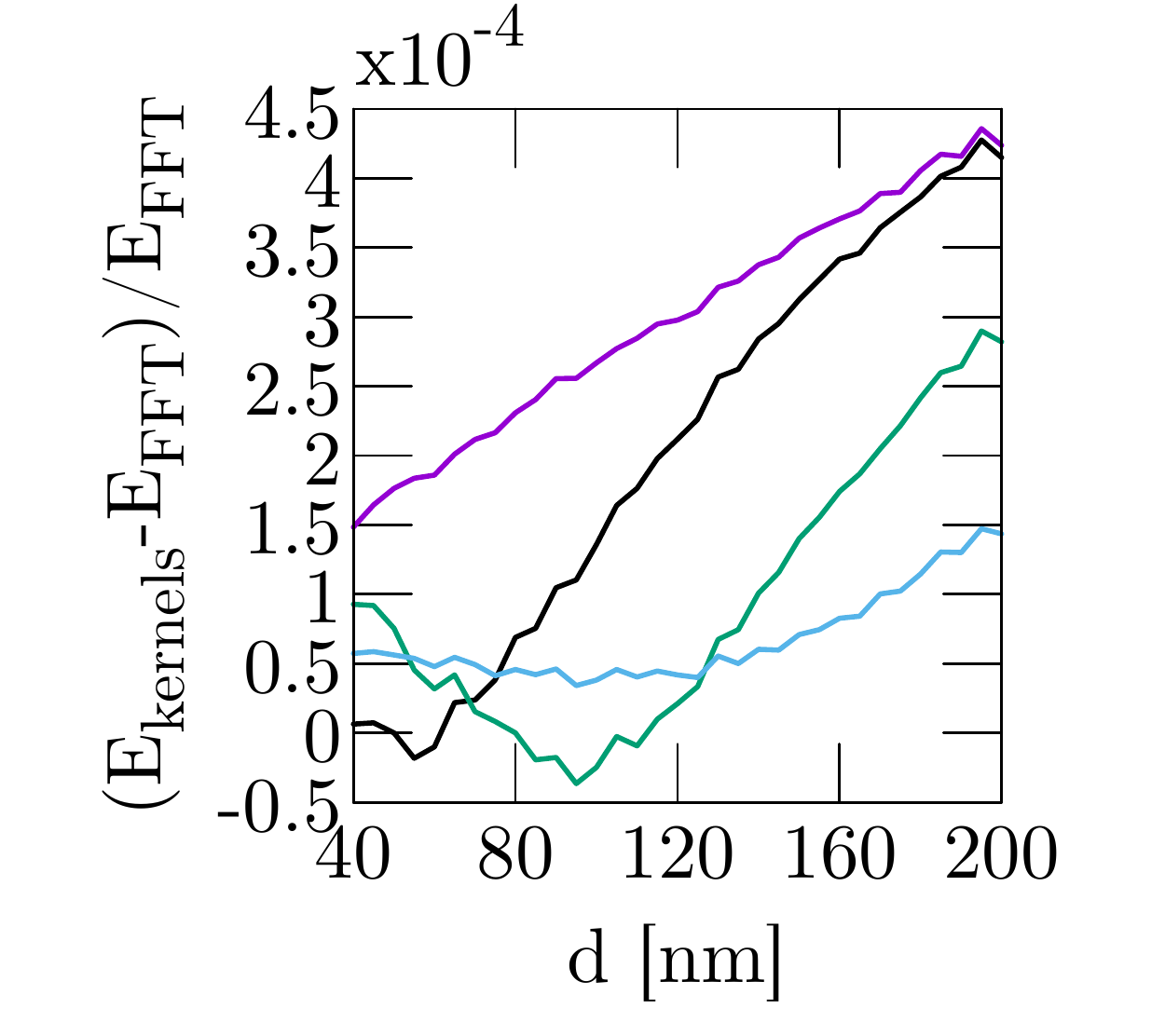}\\
\hspace{0.5cm}(c)\hspace{4cm}(d) \includegraphics[width=0.49\columnwidth]{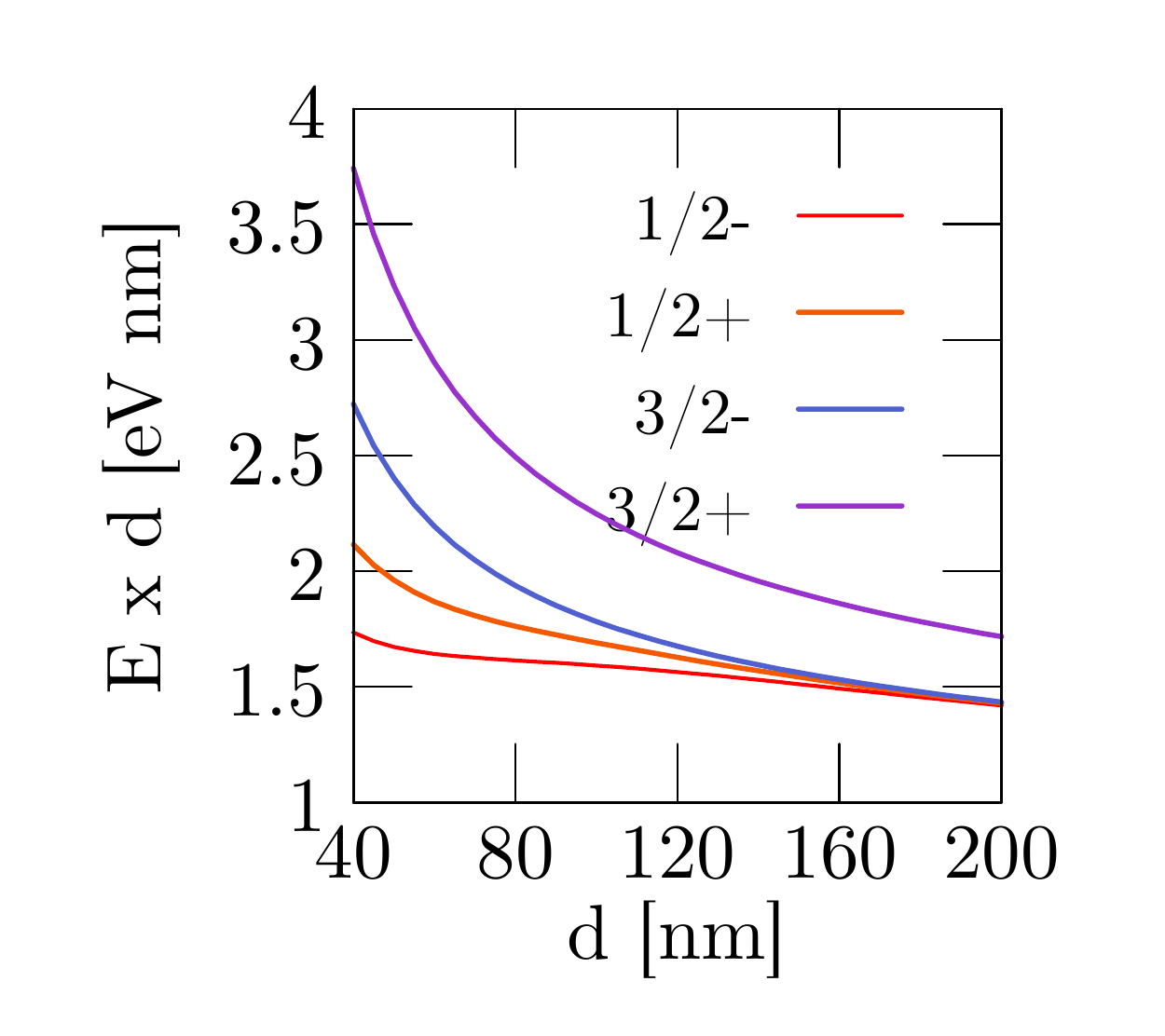}
\includegraphics[width=0.49\columnwidth]{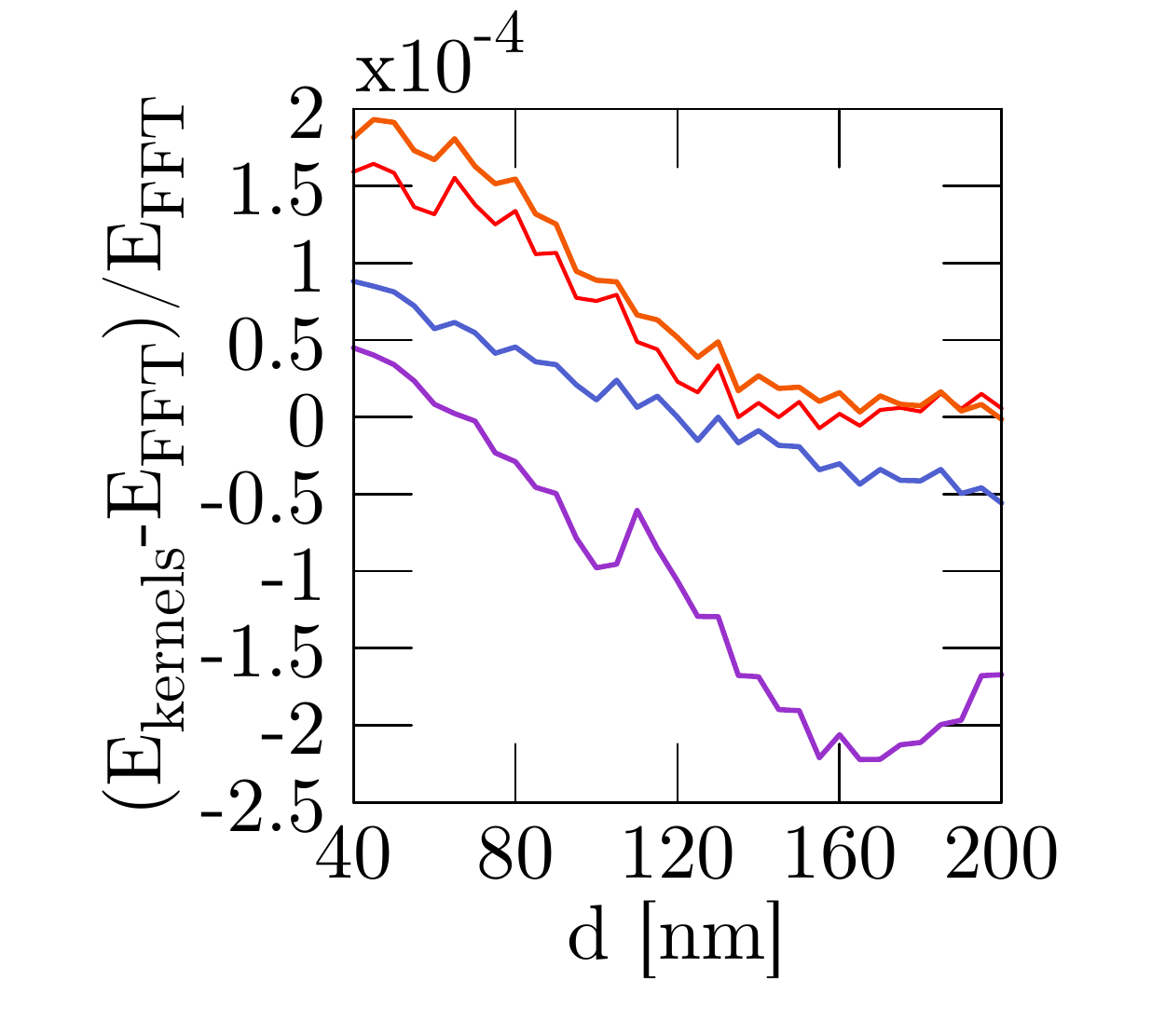}
  \caption{(a) and (c) The lowest energy levels as a function of the quantum dot length $d$ and (b) and (d) the difference between the results of both methods relative to the method (i) result for the respective energy levels shown in the same line color as in (a) and (c). The results are obtained for two electrons in (a) and (b), and for 3 electrons in (c) and (d).
  } \label{fig:1d_fun_d}
\end{figure}

For a few electrons confined in such quasi-one-dimensional systems, formation of Wigner molecules was observed \cite{Wigner1d} for sufficiently long dots.
Figures \ref{fig:1d_fun_d}(a) and \ref{fig:1d_fun_d}(c) show the energies as a function of the length $d$ of the potential well in $z$ for two and three electrons confined in the dot, respectively. The calculations are done for mesh $n_x=256$. For the evaluation of the kernels for our method we used the parameters: $S=4$, $K=65$, an initial learning rate of 0.02 decayed by $\epsilon=0.65$ in ten steps, and $N_{it}=300$ iterations.

In Figs.~\ref{fig:1d_fun_d}(b) and \ref{fig:1d_fun_d}(d) the relative difference between the results of method (ii) and method (i) is shown. The line colors correspond to the energy levels in Figs.~\ref{fig:1d_fun_d}(a) and \ref{fig:1d_fun_d}(c). The relative error is of the order of $10^{-4}$, which allows for a sufficiently good evaluation of the energy levels. 

\section{Summary}
The calculation of the energy levels of many-body quantum systems is a long-established challenge. Even with the approximate methods including DFT and CI, the computation is time-consuming due to its high complexity, resulting from the need to evaluate a large number of two-electron integrals, among other causes. 
The aim of this work was to develop a fast and efficient approach to calculate the two-electron integrals for the few-electron calculations. For many problems, it is not crucial to obtain extremely high precision of the integration, and the acceleration of the computation is beneficial provided that the error is much smaller than the order of magnitude of the energies in the system.
Our method allows us to significantly reduce the computation time, while maintaining reasonable accuracy. Picking the number of scales in our method one can choose between higher precision and faster computation.
 The optimized evaluation of the two-electron integrals can also be used in other methods used on a discrete mesh, e.g., the Hartree-Fock method.

\section{Acknowledgment}
 This research was supported in part by PLGrid Infrastructure.

\appendix

\section{Shooting method}
\label{shooting} 
The problem of two electrons confined in a 2D harmonic potential can be solved semi-analytically \cite{SZAFRAN1999}. The relative motion of the electrons is described in cylindrical coordinates by the Hamiltonian (\ref{eq:dh_CM_polar}). We solve it in a discrete mesh using the shooting method.
The wavefunction of the relative motion of the electrons $\psi(\rho, \phi) = R(\rho)e^{iM\phi}$, and the mesh is discretized into nodes $\rho_i$.
The relative motion of electrons written in cylindrical coordinates
\begin{equation}
\label{eq:h_shoot}
-\frac{1}{2} \left( \frac{d^2}{d\rho^2} +\frac{1}{\rho}\frac{d}{d\rho} - \frac{M^2}{\rho^2} - \frac{\gamma_x^2}{4}\rho^2-\frac{1}{\rho} \right) R(\rho) = \frac{E}{4} R(\rho), 
\end{equation}
can be written in the finite-difference approximation with $E'=E/4$,
\begin{eqnarray}
\label{eq:shooting}
\left( \frac{1}{(\Delta\rho)^2} + \frac{1}{2 \rho  \Delta\rho} \right) R_{i+1}  \nonumber \\
=\left( \frac{2}{(\Delta\rho)^2} + \frac{M}{\rho^2} + \frac{\gamma_x^2}{4} \rho^2 + \frac{1}{\rho} - 2E' \right) R_{i} \\
+ \left(- \frac{1}{(\Delta\rho)^2} + \frac{1}{2 \rho  \Delta\rho} \right) R_{i-1}, \nonumber 
\end{eqnarray}
where $R_{i}=R(\rho_i)$ is the wave function at node $\rho_i$ of the finite-difference mesh. In the shooting method we assume the boundary condition $R_0=0$ at the left edge of the mesh, and for a given energy we calculate the values of $R_i$ at the nodes of the mesh. We proceed to the right edge of the mesh, and $R_{n_x}$ needs to vanish. This condition is satisfied at discrete values of energy. The problem essentially is to find energies $E'$ at which $R_{n_x}=0$ in Eq.~(\ref{eq:shooting}).
 
\section{Effectiveness of the method}
 
\begin{figure}
\hspace{0.cm}(a)\hspace{2.4cm}(b) \hspace{2.4cm}(c)
 \includegraphics[width=0.32\columnwidth,trim={0.6cm 0 1.2cm 0},clip]{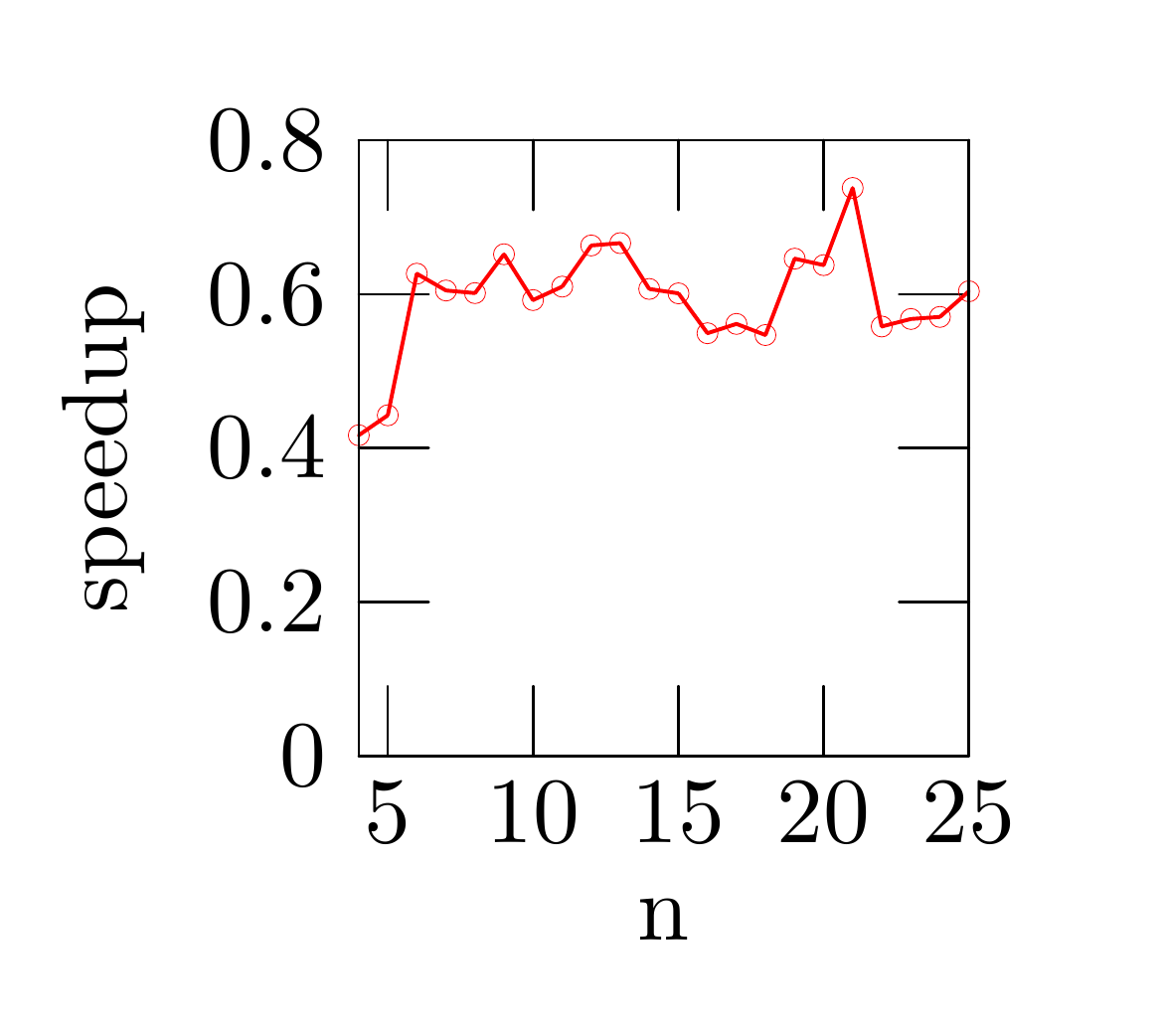}
 \includegraphics[width=0.32\columnwidth,trim={0.6cm 0 1.2cm 0},clip]{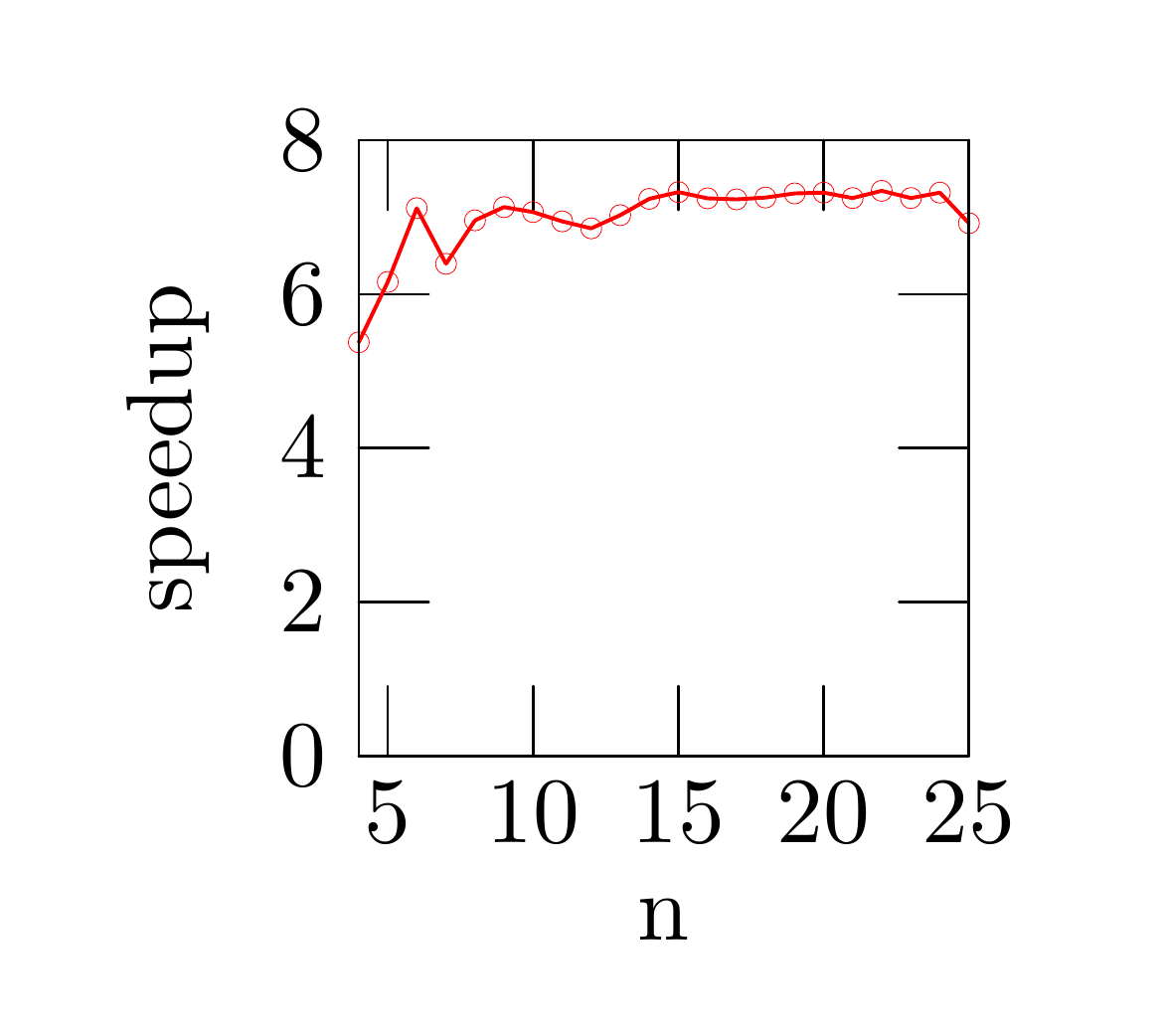}
 \includegraphics[width=0.33\columnwidth,trim={0.4cm 0 1.2cm 0},clip]{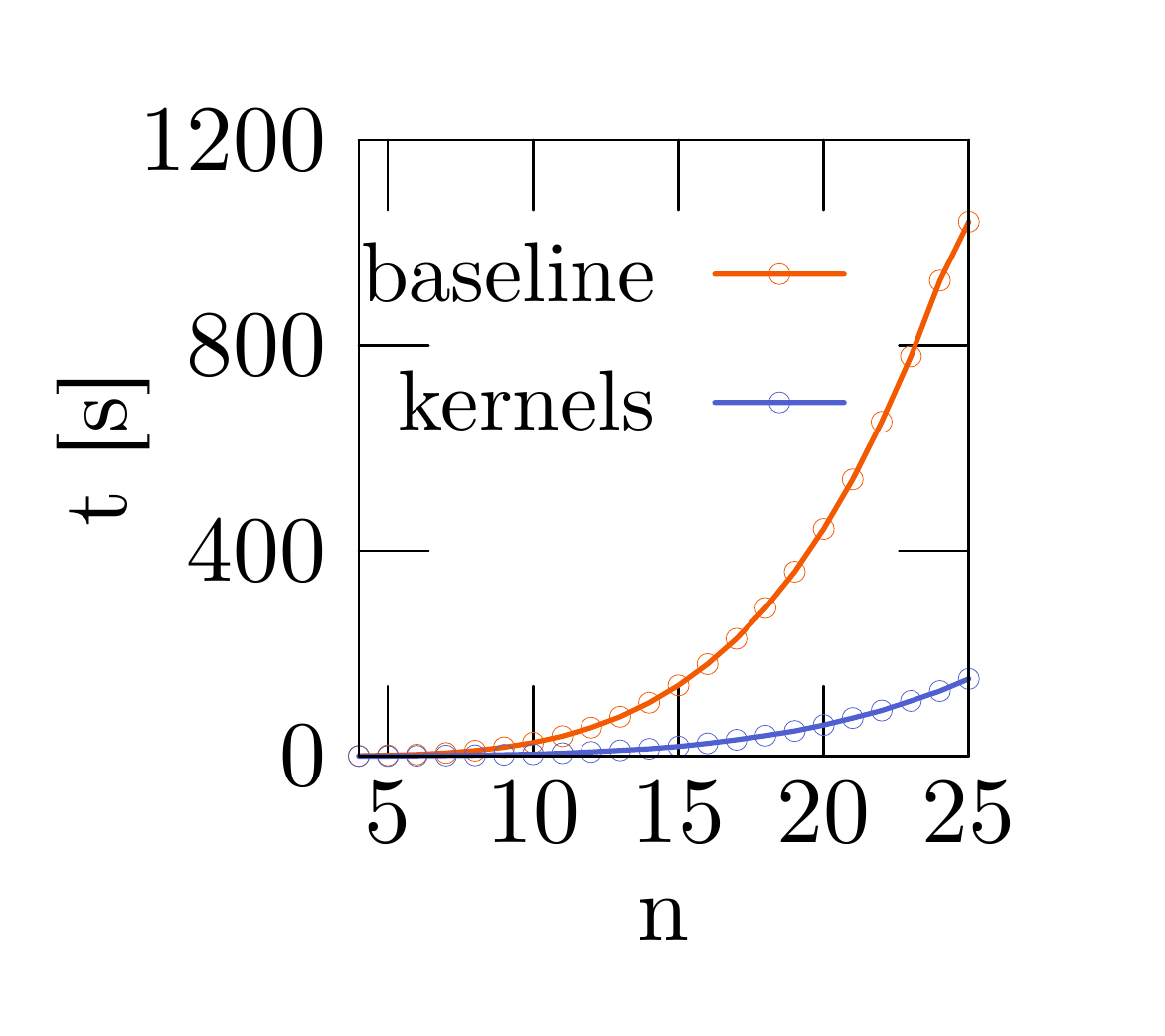}
  \caption{Speedup of the method as a function of the number of base states $n$ for calculations of two electrons in (a) one dimension and (b) two dimensions and (c) the time of the calculation for the 2D case. 
  } \label{fig:speed}
\end{figure}

We consider the effectiveness of the method with respect to the number of electrons and dimensionality. 
We use the CI method, with $n$ basis states $\phi_k(\mathbf{r})$ from which we form the Slater determinants [Eq.~(\ref{eq:Slater})], and the number of the two-electron integrals depends only on the size of the basis $n$, irrespective of the number of electrons. In our method the calculation of two-electron integrals is optimized via ML, thus the speedup depends only on the number of the basis states. In Fig.~\ref{fig:speed} we present the speedup (the ratio of the time of calculation by our method to the baseline time) as a function of $n$, obtained for electrons in one dimension [Fig.~\ref{fig:speed}(a)] and two dimensions [Fig.~\ref{fig:speed}(b)] on a mesh with $n_x=256$, and a kernel with $K=65$, $S=4$. In two dimensions the calculation with our method is several times faster than using FFT.
In one dimension our method is slightly slower than the baseline because it performs several additional operations (upsampling, downsampling) that in one dimension get less boost in parallelization. Importantly, our method gains more boost for two or more dimensions, and the results in one dimension are shown for presentation purposes.

Table \ref{tab:precision_int} shows the total energies and the interaction energies of the two lowest levels  
calculated for two to four electrons and the error of our method relative to the baseline. The error tends to increase with the number of electrons; however, it does not change linearly, as one would expect. We found that the increase is of the same order as the increase of the interaction energy. The reason is that our method optimizes the evaluation of the interaction integrals; thus, the error will scale in a manner similar to the interaction energy.

\begin{table}[htbp]
\begin{center}
\begin{tabular}{|c|c|c|c|c|}
\hline
$N$   &	Level	& $E_{FFT}$	(eV)	& $E_{int}$	(eV) & $E_{FFT}-E_{kernels}$	(eV)	\\
\hline
  2	&	1	&	0.32727	&	0.04727	&	$6\times 10^{-6}$ \\
  2	&	2	&	0.44457	&	0.16457	&	$3.2\times 10^{-5}$ \\
\hline
  3	&	1	&	0.66788	&	0.24788	&	$6.1\times 10^{-5}$ \\ 
  3	&	2	&	0.773875&	0.353875&	$9.6\times 10^{-5}$ \\
\hline
  4	&	1	&	1.03282 	&	0.47282 	&	$1.6\times 10^{-4}$ \\
  4	&	2	&	1.04097	&	0.48097	&	$1.8\times 10^{-4}$ \\
\hline
\end{tabular}
\end{center}
\caption{Total energy and interaction energy of the lowest levels, and the error of our method relative to the baseline calculated for two, three and four electrons confined in a 2D harmonic potential. }
\label{tab:precision_int}
\end{table}

\bibliography{approxkernel}

\begin{thebibliography}{36}%
\makeatletter
\providecommand \@ifxundefined [1]{%
 \@ifx{#1\undefined}
}%
\providecommand \@ifnum [1]{%
 \ifnum #1\expandafter \@firstoftwo
 \else \expandafter \@secondoftwo
 \fi
}%
\providecommand \@ifx [1]{%
 \ifx #1\expandafter \@firstoftwo
 \else \expandafter \@secondoftwo
 \fi
}%
\providecommand \natexlab [1]{#1}%
\providecommand \enquote  [1]{``#1''}%
\providecommand \bibnamefont  [1]{#1}%
\providecommand \bibfnamefont [1]{#1}%
\providecommand \citenamefont [1]{#1}%
\providecommand \href@noop [0]{\@secondoftwo}%
\providecommand \href [0]{\begingroup \@sanitize@url \@href}%
\providecommand \@href[1]{\@@startlink{#1}\@@href}%
\providecommand \@@href[1]{\endgroup#1\@@endlink}%
\providecommand \@sanitize@url [0]{\catcode `\\12\catcode `\$12\catcode
  `\&12\catcode `\#12\catcode `\^12\catcode `\_12\catcode `\%12\relax}%
\providecommand \@@startlink[1]{}%
\providecommand \@@endlink[0]{}%
\providecommand \url  [0]{\begingroup\@sanitize@url \@url }%
\providecommand \@url [1]{\endgroup\@href {#1}{\urlprefix }}%
\providecommand \urlprefix  [0]{URL }%
\providecommand \Eprint [0]{\href }%
\providecommand \doibase [0]{http://dx.doi.org/}%
\providecommand \selectlanguage [0]{\@gobble}%
\providecommand \bibinfo  [0]{\@secondoftwo}%
\providecommand \bibfield  [0]{\@secondoftwo}%
\providecommand \translation [1]{[#1]}%
\providecommand \BibitemOpen [0]{}%
\providecommand \bibitemStop [0]{}%
\providecommand \bibitemNoStop [0]{.\EOS\space}%
\providecommand \EOS [0]{\spacefactor3000\relax}%
\providecommand \BibitemShut  [1]{\csname bibitem#1\endcsname}%
\let\auto@bib@innerbib\@empty
\bibitem [{\citenamefont {Carleo}\ and\ \citenamefont
  {Troyer}(2017)}]{Carleo2017}%
  \BibitemOpen
  \bibfield  {author} {\bibinfo {author} {\bibfnamefont {Giuseppe}\
  \bibnamefont {Carleo}}\ and\ \bibinfo {author} {\bibfnamefont {Matthias}\
  \bibnamefont {Troyer}},\ }\bibfield  {title} {\enquote {\bibinfo {title}
  {Solving the quantum many-body problem with artificial neural networks},}\
  }\href {\doibase 10.1126/science.aag2302} {\bibfield  {journal} {\bibinfo
  {journal} {Science}\ }\textbf {\bibinfo {volume} {355}},\ \bibinfo {pages}
  {602} (\bibinfo {year} {2017})}\BibitemShut {NoStop}%
\bibitem [{\citenamefont {Carleo}\ \emph {et~al.}(2018)\citenamefont {Carleo},
  \citenamefont {Nomura},\ and\ \citenamefont {Imada}}]{Carleo2018}%
  \BibitemOpen
  \bibfield  {author} {\bibinfo {author} {\bibfnamefont {Giuseppe}\
  \bibnamefont {Carleo}}, \bibinfo {author} {\bibfnamefont {Yusuke}\
  \bibnamefont {Nomura}}, \ and\ \bibinfo {author} {\bibfnamefont {Masatoshi}\
  \bibnamefont {Imada}},\ }\bibfield  {title} {\enquote {\bibinfo {title}
  {Constructing exact representations of quantum many-body systems with deep
  neural networks},}\ }\href {\doibase 10.1038/s41467-018-07520-3} {\bibfield
  {journal} {\bibinfo  {journal} {Nature Communications}\ }\textbf {\bibinfo
  {volume} {9}},\ \bibinfo {pages} {5322} (\bibinfo {year} {2018})}\BibitemShut
  {NoStop}%
\bibitem [{\citenamefont {Gardas}\ \emph {et~al.}(2018)\citenamefont {Gardas},
  \citenamefont {Rams},\ and\ \citenamefont {Dziarmaga}}]{Gardas2018}%
  \BibitemOpen
  \bibfield  {author} {\bibinfo {author} {\bibfnamefont {Bart\l{}omiej}\
  \bibnamefont {Gardas}}, \bibinfo {author} {\bibfnamefont {Marek~M.}\
  \bibnamefont {Rams}}, \ and\ \bibinfo {author} {\bibfnamefont {Jacek}\
  \bibnamefont {Dziarmaga}},\ }\bibfield  {title} {\enquote {\bibinfo {title}
  {Quantum neural networks to simulate many-body quantum systems},}\ }\href
  {\doibase 10.1103/PhysRevB.98.184304} {\bibfield  {journal} {\bibinfo
  {journal} {Phys. Rev. B}\ }\textbf {\bibinfo {volume} {98}},\ \bibinfo
  {pages} {184304} (\bibinfo {year} {2018})}\BibitemShut {NoStop}%
\bibitem [{\citenamefont {Cai}\ and\ \citenamefont {Liu}(2018)}]{Cai2018}%
  \BibitemOpen
  \bibfield  {author} {\bibinfo {author} {\bibfnamefont {Zi}~\bibnamefont
  {Cai}}\ and\ \bibinfo {author} {\bibfnamefont {Jinguo}\ \bibnamefont {Liu}},\
  }\bibfield  {title} {\enquote {\bibinfo {title} {Approximating quantum
  many-body wave functions using artificial neural networks},}\ }\href
  {\doibase 10.1103/PhysRevB.97.035116} {\bibfield  {journal} {\bibinfo
  {journal} {Phys. Rev. B}\ }\textbf {\bibinfo {volume} {97}},\ \bibinfo
  {pages} {035116} (\bibinfo {year} {2018})}\BibitemShut {NoStop}%
\bibitem [{\citenamefont {Fabrizio}\ \emph {et~al.}(2019)\citenamefont
  {Fabrizio}, \citenamefont {Grisafi}, \citenamefont {Meyer}, \citenamefont
  {Ceriotti},\ and\ \citenamefont {Corminboeuf}}]{Fabrizio2019}%
  \BibitemOpen
  \bibfield  {author} {\bibinfo {author} {\bibfnamefont {Alberto}\ \bibnamefont
  {Fabrizio}}, \bibinfo {author} {\bibfnamefont {Andrea}\ \bibnamefont
  {Grisafi}}, \bibinfo {author} {\bibfnamefont {Benjamin}\ \bibnamefont
  {Meyer}}, \bibinfo {author} {\bibfnamefont {Michele}\ \bibnamefont
  {Ceriotti}}, \ and\ \bibinfo {author} {\bibfnamefont {Clemence}\ \bibnamefont
  {Corminboeuf}},\ }\bibfield  {title} {\enquote {\bibinfo {title} {Electron
  density learning of non-covalent systems},}\ }\href {\doibase
  10.1039/C9SC02696G} {\bibfield  {journal} {\bibinfo  {journal} {Chem. Sci.}\
  }\textbf {\bibinfo {volume} {10}},\ \bibinfo {pages} {9424} (\bibinfo {year}
  {2019})}\BibitemShut {NoStop}%
\bibitem [{\citenamefont {Sch{\"u}tt}\ \emph {et~al.}(2019)\citenamefont
  {Sch{\"u}tt}, \citenamefont {Gastegger}, \citenamefont {Tkatchenko},
  \citenamefont {M{\"u}ller},\ and\ \citenamefont {Maurer}}]{Schutt2019}%
  \BibitemOpen
  \bibfield  {author} {\bibinfo {author} {\bibfnamefont {K.~T.}\ \bibnamefont
  {Sch{\"u}tt}}, \bibinfo {author} {\bibfnamefont {M.}~\bibnamefont
  {Gastegger}}, \bibinfo {author} {\bibfnamefont {A.}~\bibnamefont
  {Tkatchenko}}, \bibinfo {author} {\bibfnamefont {K.-R.}\ \bibnamefont
  {M{\"u}ller}}, \ and\ \bibinfo {author} {\bibfnamefont {R.~J.}\ \bibnamefont
  {Maurer}},\ }\bibfield  {title} {\enquote {\bibinfo {title} {Unifying machine
  learning and quantum chemistry with a deep neural network for molecular
  wavefunctions},}\ }\href {\doibase 10.1038/s41467-019-12875-2} {\bibfield
  {journal} {\bibinfo  {journal} {Nat. Commun.}\ }\textbf {\bibinfo {volume}
  {10}},\ \bibinfo {pages} {5024} (\bibinfo {year} {2019})}\BibitemShut
  {NoStop}%
\bibitem [{\citenamefont {Grisafi}\ \emph {et~al.}(2019)\citenamefont
  {Grisafi}, \citenamefont {Fabrizio}, \citenamefont {Meyer}, \citenamefont
  {Wilkins}, \citenamefont {Corminboeuf},\ and\ \citenamefont
  {Ceriotti}}]{Grisafi2019}%
  \BibitemOpen
  \bibfield  {author} {\bibinfo {author} {\bibfnamefont {Andrea}\ \bibnamefont
  {Grisafi}}, \bibinfo {author} {\bibfnamefont {Alberto}\ \bibnamefont
  {Fabrizio}}, \bibinfo {author} {\bibfnamefont {Benjamin}\ \bibnamefont
  {Meyer}}, \bibinfo {author} {\bibfnamefont {David~M.}\ \bibnamefont
  {Wilkins}}, \bibinfo {author} {\bibfnamefont {Clemence}\ \bibnamefont
  {Corminboeuf}}, \ and\ \bibinfo {author} {\bibfnamefont {Michele}\
  \bibnamefont {Ceriotti}},\ }\bibfield  {title} {\enquote {\bibinfo {title}
  {Transferable machine-learning model of the electron density},}\ }\href
  {\doibase 10.1021/acscentsci.8b00551} {\bibfield  {journal} {\bibinfo
  {journal} {ACS Cent. Sci.}\ }\textbf {\bibinfo {volume} {5}},\ \bibinfo
  {pages} {57} (\bibinfo {year} {2019})}\BibitemShut {NoStop}%
\bibitem [{\citenamefont {Han}\ \emph {et~al.}(2019)\citenamefont {Han},
  \citenamefont {Zhang},\ and\ \citenamefont {E}}]{Han2019}%
  \BibitemOpen
  \bibfield  {author} {\bibinfo {author} {\bibfnamefont {Jiequn}\ \bibnamefont
  {Han}}, \bibinfo {author} {\bibfnamefont {Linfeng}\ \bibnamefont {Zhang}}, \
  and\ \bibinfo {author} {\bibfnamefont {Weinan}\ \bibnamefont {E}},\
  }\bibfield  {title} {\enquote {\bibinfo {title} {Solving many-electron
  {Schr\"odinger} equation using deep neural networks},}\ }\href {\doibase
  https://doi.org/10.1016/j.jcp.2019.108929} {\bibfield  {journal} {\bibinfo
  {journal} {J. Comput. Phys.}\ }\textbf {\bibinfo {volume} {399}},\ \bibinfo
  {pages} {108929} (\bibinfo {year} {2019})}\BibitemShut {NoStop}%
\bibitem [{\citenamefont {Hermann}\ \emph {et~al.}()\citenamefont {Hermann},
  \citenamefont {Sch\"atzle},\ and\ \citenamefont {No\'e}}]{Hermann2019}%
  \BibitemOpen
  \bibfield  {author} {\bibinfo {author} {\bibfnamefont {Jan}\ \bibnamefont
  {Hermann}}, \bibinfo {author} {\bibfnamefont {Zeno}\ \bibnamefont
  {Sch\"atzle}}, \ and\ \bibinfo {author} {\bibfnamefont {Frank}\ \bibnamefont
  {No\'e}},\ }\bibfield  {title} {\enquote {\bibinfo {title} {Deep neural
  network solution of the electronic {Schr\"odinger} equation},}\ }\href@noop
  {} {\ }\BibitemShut {NoStop}%
\bibitem [{\citenamefont {Rigo}\ and\ \citenamefont
  {Mitchell}(2020)}]{Rigo2020}%
  \BibitemOpen
  \bibfield  {author} {\bibinfo {author} {\bibfnamefont {Jonas~B.}\
  \bibnamefont {Rigo}}\ and\ \bibinfo {author} {\bibfnamefont {Andrew~K.}\
  \bibnamefont {Mitchell}},\ }\bibfield  {title} {\enquote {\bibinfo {title}
  {Machine learning effective models for quantum systems},}\ }\href {\doibase
  10.1103/PhysRevB.101.241105} {\bibfield  {journal} {\bibinfo  {journal}
  {Phys. Rev. B}\ }\textbf {\bibinfo {volume} {101}},\ \bibinfo {pages}
  {241105(R)} (\bibinfo {year} {2020})}\BibitemShut {NoStop}%
\bibitem [{\citenamefont {Liu}\ \emph {et~al.}(2017)\citenamefont {Liu},
  \citenamefont {Qi}, \citenamefont {Meng},\ and\ \citenamefont
  {Fu}}]{Liu2017}%
  \BibitemOpen
  \bibfield  {author} {\bibinfo {author} {\bibfnamefont {Junwei}\ \bibnamefont
  {Liu}}, \bibinfo {author} {\bibfnamefont {Yang}\ \bibnamefont {Qi}}, \bibinfo
  {author} {\bibfnamefont {Zi~Yang}\ \bibnamefont {Meng}}, \ and\ \bibinfo
  {author} {\bibfnamefont {Liang}\ \bibnamefont {Fu}},\ }\bibfield  {title}
  {\enquote {\bibinfo {title} {Self-learning {Monte Carlo} method},}\ }\href
  {\doibase 10.1103/PhysRevB.95.041101} {\bibfield  {journal} {\bibinfo
  {journal} {Phys. Rev. B}\ }\textbf {\bibinfo {volume} {95}},\ \bibinfo
  {pages} {041101(R)} (\bibinfo {year} {2017})}\BibitemShut {NoStop}%
\bibitem [{\citenamefont {Nagai}\ \emph {et~al.}(2017)\citenamefont {Nagai},
  \citenamefont {Shen}, \citenamefont {Qi}, \citenamefont {Liu},\ and\
  \citenamefont {Fu}}]{Nagai2017}%
  \BibitemOpen
  \bibfield  {author} {\bibinfo {author} {\bibfnamefont {Yuki}\ \bibnamefont
  {Nagai}}, \bibinfo {author} {\bibfnamefont {Huitao}\ \bibnamefont {Shen}},
  \bibinfo {author} {\bibfnamefont {Yang}\ \bibnamefont {Qi}}, \bibinfo
  {author} {\bibfnamefont {Junwei}\ \bibnamefont {Liu}}, \ and\ \bibinfo
  {author} {\bibfnamefont {Liang}\ \bibnamefont {Fu}},\ }\bibfield  {title}
  {\enquote {\bibinfo {title} {Self-learning {Monte Carlo} method:
  Continuous-time algorithm},}\ }\href {\doibase 10.1103/PhysRevB.96.161102}
  {\bibfield  {journal} {\bibinfo  {journal} {Phys. Rev. B}\ }\textbf {\bibinfo
  {volume} {96}},\ \bibinfo {pages} {161102(R)} (\bibinfo {year}
  {2017})}\BibitemShut {NoStop}%
\bibitem [{\citenamefont {Shen}\ \emph {et~al.}(2018)\citenamefont {Shen},
  \citenamefont {Liu},\ and\ \citenamefont {Fu}}]{Shen2018}%
  \BibitemOpen
  \bibfield  {author} {\bibinfo {author} {\bibfnamefont {Huitao}\ \bibnamefont
  {Shen}}, \bibinfo {author} {\bibfnamefont {Junwei}\ \bibnamefont {Liu}}, \
  and\ \bibinfo {author} {\bibfnamefont {Liang}\ \bibnamefont {Fu}},\
  }\bibfield  {title} {\enquote {\bibinfo {title} {Self-learning {Monte Carlo}
  with deep neural networks},}\ }\href {\doibase 10.1103/PhysRevB.97.205140}
  {\bibfield  {journal} {\bibinfo  {journal} {Phys. Rev. B}\ }\textbf {\bibinfo
  {volume} {97}},\ \bibinfo {pages} {205140} (\bibinfo {year}
  {2018})}\BibitemShut {NoStop}%
\bibitem [{\citenamefont {Hohenberg}\ and\ \citenamefont
  {Kohn}(1964)}]{Hohenberg1964}%
  \BibitemOpen
  \bibfield  {author} {\bibinfo {author} {\bibfnamefont {P.}~\bibnamefont
  {Hohenberg}}\ and\ \bibinfo {author} {\bibfnamefont {W.}~\bibnamefont
  {Kohn}},\ }\bibfield  {title} {\enquote {\bibinfo {title} {Inhomogeneous
  electron gas},}\ }\href {\doibase 10.1103/PhysRev.136.B864} {\bibfield
  {journal} {\bibinfo  {journal} {Phys. Rev.}\ }\textbf {\bibinfo {volume}
  {136}},\ \bibinfo {pages} {B864} (\bibinfo {year} {1964})}\BibitemShut
  {NoStop}%
\bibitem [{\citenamefont {Hartree}\ and\ \citenamefont
  {Hartree}(1935)}]{Hartree1935}%
  \BibitemOpen
  \bibfield  {author} {\bibinfo {author} {\bibfnamefont {Douglas~Rayner}\
  \bibnamefont {Hartree}}\ and\ \bibinfo {author} {\bibfnamefont
  {W.}~\bibnamefont {Hartree}},\ }\bibfield  {title} {\enquote {\bibinfo
  {title} {Self-consistent field, with exchange, for beryllium},}\ }\href
  {\doibase 10.1098/rspa.1935.0085} {\bibfield  {journal} {\bibinfo  {journal}
  {Proc. R. Soc. London, Ser A.}\ }\textbf {\bibinfo {volume} {150}},\ \bibinfo
  {pages} {9} (\bibinfo {year} {1935})}\BibitemShut {NoStop}%
\bibitem [{\citenamefont {Roothaan}(1960)}]{Roothaan1960}%
  \BibitemOpen
  \bibfield  {author} {\bibinfo {author} {\bibfnamefont {C.~C.~J.}\
  \bibnamefont {Roothaan}},\ }\bibfield  {title} {\enquote {\bibinfo {title}
  {Self-consistent field theory for open shells of electronic systems},}\
  }\href {\doibase 10.1103/RevModPhys.32.179} {\bibfield  {journal} {\bibinfo
  {journal} {Rev. Mod. Phys.}\ }\textbf {\bibinfo {volume} {32}},\ \bibinfo
  {pages} {179} (\bibinfo {year} {1960})}\BibitemShut {NoStop}%
\bibitem [{\citenamefont {Shavitt}(1977)}]{Shavitt1977}%
  \BibitemOpen
  \bibfield  {author} {\bibinfo {author} {\bibfnamefont {Isaiah}\ \bibnamefont
  {Shavitt}},\ }\enquote {\bibinfo {title} {The method of configuration
  interaction},}\ in\ \href {\doibase 10.1007/978-1-4757-0887-5_6} {\emph
  {\bibinfo {booktitle} {Methods of Electronic Structure Theory}}},\ \bibinfo
  {editor} {edited by\ \bibinfo {editor} {\bibfnamefont {Henry~F.}\
  \bibnamefont {Schaefer}}}\ (\bibinfo  {publisher} {Springer US},\ \bibinfo
  {address} {Boston, MA},\ \bibinfo {year} {1977})\ p.\ \bibinfo {pages}
  {189}\BibitemShut {NoStop}%
\bibitem [{\citenamefont {Pople}\ \emph {et~al.}(1987)\citenamefont {Pople},
  \citenamefont {Head-Gordon},\ and\ \citenamefont {Raghavachari}}]{Pople1987}%
  \BibitemOpen
  \bibfield  {author} {\bibinfo {author} {\bibfnamefont {John~A.}\ \bibnamefont
  {Pople}}, \bibinfo {author} {\bibfnamefont {Martin}\ \bibnamefont
  {Head-Gordon}}, \ and\ \bibinfo {author} {\bibfnamefont {Krishnan}\
  \bibnamefont {Raghavachari}},\ }\bibfield  {title} {\enquote {\bibinfo
  {title} {Quadratic configuration interaction. {A} general technique for
  determining electron correlation energies},}\ }\href {\doibase
  10.1063/1.453520} {\bibfield  {journal} {\bibinfo  {journal} {J. Chem.
  Phys.}\ }\textbf {\bibinfo {volume} {87}},\ \bibinfo {pages} {5968} (\bibinfo
  {year} {1987})}\BibitemShut {NoStop}%
\bibitem [{\citenamefont {Rontani}\ \emph {et~al.}(2006)\citenamefont
  {Rontani}, \citenamefont {Cavazzoni}, \citenamefont {Bellucci},\ and\
  \citenamefont {Goldoni}}]{Rontani2006}%
  \BibitemOpen
  \bibfield  {author} {\bibinfo {author} {\bibfnamefont {Massimo}\ \bibnamefont
  {Rontani}}, \bibinfo {author} {\bibfnamefont {Carlo}\ \bibnamefont
  {Cavazzoni}}, \bibinfo {author} {\bibfnamefont {Devis}\ \bibnamefont
  {Bellucci}}, \ and\ \bibinfo {author} {\bibfnamefont {Guido}\ \bibnamefont
  {Goldoni}},\ }\bibfield  {title} {\enquote {\bibinfo {title} {Full
  configuration interaction approach to the few-electron problem in artificial
  atoms},}\ }\href {\doibase 10.1063/1.2179418} {\bibfield  {journal} {\bibinfo
   {journal} {J. Chem. Phys.}\ }\textbf {\bibinfo {volume} {124}},\ \bibinfo
  {pages} {124102} (\bibinfo {year} {2006})}\BibitemShut {NoStop}%
\bibitem [{\citenamefont {Puerto~Gimenez}\ \emph {et~al.}(2007)\citenamefont
  {Puerto~Gimenez}, \citenamefont {Korkusinski},\ and\ \citenamefont
  {Hawrylak}}]{Gimenez2007}%
  \BibitemOpen
  \bibfield  {author} {\bibinfo {author} {\bibfnamefont {Irene}\ \bibnamefont
  {Puerto~Gimenez}}, \bibinfo {author} {\bibfnamefont {Marek}\ \bibnamefont
  {Korkusinski}}, \ and\ \bibinfo {author} {\bibfnamefont {Pawel}\ \bibnamefont
  {Hawrylak}},\ }\bibfield  {title} {\enquote {\bibinfo {title} {Linear
  combination of harmonic orbitals and configuration interaction method for the
  voltage control of exchange interaction in gated lateral quantum dot
  networks},}\ }\href {\doibase 10.1103/PhysRevB.76.075336} {\bibfield
  {journal} {\bibinfo  {journal} {Phys. Rev. B}\ }\textbf {\bibinfo {volume}
  {76}},\ \bibinfo {pages} {075336} (\bibinfo {year} {2007})}\BibitemShut
  {NoStop}%
\bibitem [{\citenamefont {Beylkin}\ \emph {et~al.}(2009)\citenamefont
  {Beylkin}, \citenamefont {Kurcz},\ and\ \citenamefont
  {Monz\'on}}]{Beylkin2009}%
  \BibitemOpen
  \bibfield  {author} {\bibinfo {author} {\bibfnamefont {Gregory}\ \bibnamefont
  {Beylkin}}, \bibinfo {author} {\bibfnamefont {Christopher}\ \bibnamefont
  {Kurcz}}, \ and\ \bibinfo {author} {\bibfnamefont {Lucas}\ \bibnamefont
  {Monz\'on}},\ }\bibfield  {title} {\enquote {\bibinfo {title} {Fast
  convolution with the free space {Helmholtz} {Green's} function},}\ }\href
  {\doibase https://doi.org/10.1016/j.jcp.2008.12.027} {\bibfield  {journal}
  {\bibinfo  {journal} {J. Comput. Phys.}\ }\textbf {\bibinfo {volume} {228}},\
  \bibinfo {pages} {2770} (\bibinfo {year} {2009})}\BibitemShut {NoStop}%
\bibitem [{\citenamefont {Genovese}\ \emph {et~al.}(2006)\citenamefont
  {Genovese}, \citenamefont {Deutsch}, \citenamefont {Neelov}, \citenamefont
  {Goedecker},\ and\ \citenamefont {Beylkin}}]{Genovese2006}%
  \BibitemOpen
  \bibfield  {author} {\bibinfo {author} {\bibfnamefont {Luigi}\ \bibnamefont
  {Genovese}}, \bibinfo {author} {\bibfnamefont {Thierry}\ \bibnamefont
  {Deutsch}}, \bibinfo {author} {\bibfnamefont {Alexey}\ \bibnamefont
  {Neelov}}, \bibinfo {author} {\bibfnamefont {Stefan}\ \bibnamefont
  {Goedecker}}, \ and\ \bibinfo {author} {\bibfnamefont {Gregory}\ \bibnamefont
  {Beylkin}},\ }\bibfield  {title} {\enquote {\bibinfo {title} {Efficient
  solution of {Poisson's} equation with free boundary conditions},}\ }\href
  {\doibase 10.1063/1.2335442} {\bibfield  {journal} {\bibinfo  {journal} {J.
  Chem. Phys.}\ }\textbf {\bibinfo {volume} {125}},\ \bibinfo {pages} {074105}
  (\bibinfo {year} {2006})}\BibitemShut {NoStop}%
\bibitem [{\citenamefont {Exl}\ \emph {et~al.}(2016)\citenamefont {Exl},
  \citenamefont {Mauser},\ and\ \citenamefont {Zhang}}]{Exl2016}%
  \BibitemOpen
  \bibfield  {author} {\bibinfo {author} {\bibfnamefont {Lukas}\ \bibnamefont
  {Exl}}, \bibinfo {author} {\bibfnamefont {Norbert~J.}\ \bibnamefont
  {Mauser}}, \ and\ \bibinfo {author} {\bibfnamefont {Yong}\ \bibnamefont
  {Zhang}},\ }\bibfield  {title} {\enquote {\bibinfo {title} {Accurate and
  efficient computation of nonlocal potentials based on gaussian-sum
  approximation},}\ }\href {\doibase https://doi.org/10.1016/j.jcp.2016.09.045}
  {\bibfield  {journal} {\bibinfo  {journal} {J. Comput. Phys.}\ }\textbf
  {\bibinfo {volume} {327}},\ \bibinfo {pages} {629} (\bibinfo {year}
  {2016})}\BibitemShut {NoStop}%
\bibitem [{\citenamefont {Mukherjee}\ \emph {et~al.}(1975)\citenamefont
  {Mukherjee}, \citenamefont {Roy},\ and\ \citenamefont {Sil}}]{Mukherjee1975}%
  \BibitemOpen
  \bibfield  {author} {\bibinfo {author} {\bibfnamefont {S.~C.}\ \bibnamefont
  {Mukherjee}}, \bibinfo {author} {\bibfnamefont {K.}~\bibnamefont {Roy}}, \
  and\ \bibinfo {author} {\bibfnamefont {N.~C.}\ \bibnamefont {Sil}},\
  }\bibfield  {title} {\enquote {\bibinfo {title} {Evaluation of the {Coulomb}
  integral for scattering problems},}\ }\href {\doibase
  10.1103/PhysRevA.12.1719} {\bibfield  {journal} {\bibinfo  {journal} {Phys.
  Rev. A}\ }\textbf {\bibinfo {volume} {12}},\ \bibinfo {pages} {1719}
  (\bibinfo {year} {1975})}\BibitemShut {NoStop}%
\bibitem [{\citenamefont {Lesiuk}\ and\ \citenamefont
  {Moszynski}(2014)}]{Lesiuk2014}%
  \BibitemOpen
  \bibfield  {author} {\bibinfo {author} {\bibfnamefont {Micha\l{}}\
  \bibnamefont {Lesiuk}}\ and\ \bibinfo {author} {\bibfnamefont {Robert}\
  \bibnamefont {Moszynski}},\ }\bibfield  {title} {\enquote {\bibinfo {title}
  {Reexamination of the calculation of two-center, two-electron integrals over
  {Slater}-type orbitals. i. {Coulomb} and hybrid integrals},}\ }\href
  {\doibase 10.1103/PhysRevE.90.063318} {\bibfield  {journal} {\bibinfo
  {journal} {Phys. Rev. E}\ }\textbf {\bibinfo {volume} {90}},\ \bibinfo
  {pages} {063318} (\bibinfo {year} {2014})}\BibitemShut {NoStop}%
\bibitem [{\citenamefont {Peels}\ and\ \citenamefont
  {Knizia}(2020)}]{Peels2020}%
  \BibitemOpen
  \bibfield  {author} {\bibinfo {author} {\bibfnamefont {Mieke}\ \bibnamefont
  {Peels}}\ and\ \bibinfo {author} {\bibfnamefont {Gerald}\ \bibnamefont
  {Knizia}},\ }\bibfield  {title} {\enquote {\bibinfo {title} {Fast evaluation
  of two-center integrals over gaussian charge distributions and gaussian
  orbitals with general interaction kernels},}\ }\href {\doibase
  10.1021/acs.jctc.9b01296} {\bibfield  {journal} {\bibinfo  {journal} {Journal
  of Chemical Theory and Computation}\ }\textbf {\bibinfo {volume} {16}},\
  \bibinfo {pages} {2570} (\bibinfo {year} {2020})}\BibitemShut {NoStop}%
\bibitem [{\citenamefont {Goodfellow}\ \emph {et~al.}(2016)\citenamefont
  {Goodfellow}, \citenamefont {Bengio},\ and\ \citenamefont
  {Courville}}]{Goodfellow-et-al-2016}%
  \BibitemOpen
  \bibfield  {author} {\bibinfo {author} {\bibfnamefont {Ian}\ \bibnamefont
  {Goodfellow}}, \bibinfo {author} {\bibfnamefont {Yoshua}\ \bibnamefont
  {Bengio}}, \ and\ \bibinfo {author} {\bibfnamefont {Aaron}\ \bibnamefont
  {Courville}},\ }\href@noop {} {\emph {\bibinfo {title} {Deep Learning}}}\
  (\bibinfo  {publisher} {MIT Press, Cambridge, MA},\ \bibinfo {year}
  {2016})\BibitemShut {NoStop}%
\bibitem [{Krz()}]{KrzysialkeGithub}%
  \BibitemOpen
  \href@noop {} {}\bibinfo {howpublished} {The source code for the
  implementation of the proposed method can be found at:
  \href{https://github.com/kmkolasinski/approxkernel}{https://github.com/kmkolasinski/approxkernel}}\BibitemShut
  {NoStop}%
\bibitem [{\citenamefont {Cudazzo}\ \emph {et~al.}(2011)\citenamefont
  {Cudazzo}, \citenamefont {Tokatly},\ and\ \citenamefont
  {Rubio}}]{Cudazzo2011}%
  \BibitemOpen
  \bibfield  {author} {\bibinfo {author} {\bibfnamefont {Pierluigi}\
  \bibnamefont {Cudazzo}}, \bibinfo {author} {\bibfnamefont {Ilya~V.}\
  \bibnamefont {Tokatly}}, \ and\ \bibinfo {author} {\bibfnamefont {Angel}\
  \bibnamefont {Rubio}},\ }\bibfield  {title} {\enquote {\bibinfo {title}
  {Dielectric screening in two-dimensional insulators: Implications for
  excitonic and impurity states in graphane},}\ }\href {\doibase
  10.1103/PhysRevB.84.085406} {\bibfield  {journal} {\bibinfo  {journal} {Phys.
  Rev. B}\ }\textbf {\bibinfo {volume} {84}},\ \bibinfo {pages} {085406}
  (\bibinfo {year} {2011})}\BibitemShut {NoStop}%
\bibitem [{\citenamefont {Bednarek}\ \emph {et~al.}(2003)\citenamefont
  {Bednarek}, \citenamefont {Szafran}, \citenamefont {Chwiej},\ and\
  \citenamefont {Adamowski}}]{1dpot}%
  \BibitemOpen
  \bibfield  {author} {\bibinfo {author} {\bibfnamefont {Stanis\l{}aw}\
  \bibnamefont {Bednarek}}, \bibinfo {author} {\bibfnamefont {Bart\l{}omiej}\
  \bibnamefont {Szafran}}, \bibinfo {author} {\bibfnamefont {Tomasz}\
  \bibnamefont {Chwiej}}, \ and\ \bibinfo {author} {\bibfnamefont {Janusz}\
  \bibnamefont {Adamowski}},\ }\bibfield  {title} {\enquote {\bibinfo {title}
  {Effective interaction for charge carriers confined in quasi-one-dimensional
  nanostructures},}\ }\href {\doibase 10.1103/PhysRevB.68.045328} {\bibfield
  {journal} {\bibinfo  {journal} {Phys. Rev. B}\ }\textbf {\bibinfo {volume}
  {68}},\ \bibinfo {pages} {045328} (\bibinfo {year} {2003})}\BibitemShut
  {NoStop}%
\bibitem [{\citenamefont {Szafran}\ \emph {et~al.}(2004)\citenamefont
  {Szafran}, \citenamefont {Peeters}, \citenamefont {Bednarek}, \citenamefont
  {Chwiej},\ and\ \citenamefont {Adamowski}}]{Wigner1d}%
  \BibitemOpen
  \bibfield  {author} {\bibinfo {author} {\bibfnamefont {Bart\l{}omiej}\
  \bibnamefont {Szafran}}, \bibinfo {author} {\bibfnamefont {Francois~M.}\
  \bibnamefont {Peeters}}, \bibinfo {author} {\bibfnamefont {Stanis\l{}aw}\
  \bibnamefont {Bednarek}}, \bibinfo {author} {\bibfnamefont {Tomasz}\
  \bibnamefont {Chwiej}}, \ and\ \bibinfo {author} {\bibfnamefont {Janusz}\
  \bibnamefont {Adamowski}},\ }\bibfield  {title} {\enquote {\bibinfo {title}
  {Spatial ordering of charge and spin in quasi-one-dimensional {Wigner}
  molecules},}\ }\href {\doibase 10.1103/PhysRevB.70.035401} {\bibfield
  {journal} {\bibinfo  {journal} {Phys. Rev. B}\ }\textbf {\bibinfo {volume}
  {70}},\ \bibinfo {pages} {035401} (\bibinfo {year} {2004})}\BibitemShut
  {NoStop}%
\bibitem [{mkl()}]{mkl}%
  \BibitemOpen
  \href@noop {} {}\bibinfo {howpublished} {For the Intel$\textsuperscript
  \textregistered$ MKL documentation we refer to
  \href{https://software.intel.com/en-us/mkl-developer-reference-c-convolution-and-correlation}{https://software.intel.com/en-us/mkl-developer-reference-c-convolution-and-correlation}}\BibitemShut
  {NoStop}%
\bibitem [{\citenamefont {Abadi}\ \emph {et~al.}()\citenamefont {Abadi},
  \citenamefont {Agarwal}, \citenamefont {Barham}, \citenamefont {Brevdo},
  \citenamefont {Chen}, \citenamefont {Citro}, \citenamefont {Corrado},
  \citenamefont {Davis}, \citenamefont {Dean}, \citenamefont {Devin},
  \citenamefont {Ghemawat}, \citenamefont {Goodfellow}, \citenamefont {Harp},
  \citenamefont {Irving}, \citenamefont {Isard}, \citenamefont {Jia},
  \citenamefont {Jozefowicz}, \citenamefont {Kaiser}, \citenamefont {Kudlur},
  \citenamefont {Levenberg}, \citenamefont {Man\'{e}}, \citenamefont {Monga},
  \citenamefont {Moore}, \citenamefont {Murray}, \citenamefont {Olah},
  \citenamefont {Schuster}, \citenamefont {Shlens}, \citenamefont {Steiner},
  \citenamefont {Sutskever}, \citenamefont {Talwar}, \citenamefont {Tucker},
  \citenamefont {Vanhoucke}, \citenamefont {Vasudevan}, \citenamefont
  {Vi\'{e}gas}, \citenamefont {Vinyals}, \citenamefont {Warden}, \citenamefont
  {Wattenberg}, \citenamefont {Wicke}, \citenamefont {Yu},\ and\ \citenamefont
  {Zheng}}]{tensorflow2015}%
  \BibitemOpen
  \bibfield  {author} {\bibinfo {author} {\bibfnamefont {Mart\'{\i}n}\
  \bibnamefont {Abadi}}, \bibinfo {author} {\bibfnamefont {Ashish}\
  \bibnamefont {Agarwal}}, \bibinfo {author} {\bibfnamefont {Paul}\
  \bibnamefont {Barham}}, \bibinfo {author} {\bibfnamefont {Eugene}\
  \bibnamefont {Brevdo}}, \bibinfo {author} {\bibfnamefont {Zhifeng}\
  \bibnamefont {Chen}}, \bibinfo {author} {\bibfnamefont {Craig}\ \bibnamefont
  {Citro}}, \bibinfo {author} {\bibfnamefont {Greg~S.}\ \bibnamefont
  {Corrado}}, \bibinfo {author} {\bibfnamefont {Andy}\ \bibnamefont {Davis}},
  \bibinfo {author} {\bibfnamefont {Jeffrey}\ \bibnamefont {Dean}}, \bibinfo
  {author} {\bibfnamefont {Matthieu}\ \bibnamefont {Devin}}, \bibinfo {author}
  {\bibfnamefont {Sanjay}\ \bibnamefont {Ghemawat}}, \bibinfo {author}
  {\bibfnamefont {Ian}\ \bibnamefont {Goodfellow}}, \bibinfo {author}
  {\bibfnamefont {Andrew}\ \bibnamefont {Harp}}, \bibinfo {author}
  {\bibfnamefont {Geoffrey}\ \bibnamefont {Irving}}, \bibinfo {author}
  {\bibfnamefont {Michael}\ \bibnamefont {Isard}}, \bibinfo {author}
  {\bibfnamefont {Yangqing}\ \bibnamefont {Jia}}, \bibinfo {author}
  {\bibfnamefont {Rafal}\ \bibnamefont {Jozefowicz}}, \bibinfo {author}
  {\bibfnamefont {Lukasz}\ \bibnamefont {Kaiser}}, \bibinfo {author}
  {\bibfnamefont {Manjunath}\ \bibnamefont {Kudlur}}, \bibinfo {author}
  {\bibfnamefont {Josh}\ \bibnamefont {Levenberg}}, \bibinfo {author}
  {\bibfnamefont {Dan}\ \bibnamefont {Man\'{e}}}, \bibinfo {author}
  {\bibfnamefont {Rajat}\ \bibnamefont {Monga}}, \bibinfo {author}
  {\bibfnamefont {Sherry}\ \bibnamefont {Moore}}, \bibinfo {author}
  {\bibfnamefont {Derek}\ \bibnamefont {Murray}}, \bibinfo {author}
  {\bibfnamefont {Chris}\ \bibnamefont {Olah}}, \bibinfo {author}
  {\bibfnamefont {Mike}\ \bibnamefont {Schuster}}, \bibinfo {author}
  {\bibfnamefont {Jonathon}\ \bibnamefont {Shlens}}, \bibinfo {author}
  {\bibfnamefont {Benoit}\ \bibnamefont {Steiner}}, \bibinfo {author}
  {\bibfnamefont {Ilya}\ \bibnamefont {Sutskever}}, \bibinfo {author}
  {\bibfnamefont {Kunal}\ \bibnamefont {Talwar}}, \bibinfo {author}
  {\bibfnamefont {Paul}\ \bibnamefont {Tucker}}, \bibinfo {author}
  {\bibfnamefont {Vincent}\ \bibnamefont {Vanhoucke}}, \bibinfo {author}
  {\bibfnamefont {Vijay}\ \bibnamefont {Vasudevan}}, \bibinfo {author}
  {\bibfnamefont {Fernanda}\ \bibnamefont {Vi\'{e}gas}}, \bibinfo {author}
  {\bibfnamefont {Oriol}\ \bibnamefont {Vinyals}}, \bibinfo {author}
  {\bibfnamefont {Pete}\ \bibnamefont {Warden}}, \bibinfo {author}
  {\bibfnamefont {Martin}\ \bibnamefont {Wattenberg}}, \bibinfo {author}
  {\bibfnamefont {Martin}\ \bibnamefont {Wicke}}, \bibinfo {author}
  {\bibfnamefont {Yuan}\ \bibnamefont {Yu}}, \ and\ \bibinfo {author}
  {\bibfnamefont {Xiaoqiang}\ \bibnamefont {Zheng}},\ }\href
  {http://tensorflow.org/} {\enquote {\bibinfo {title} {{TensorFlow}:
  Large-scale machine learning on heterogeneous systems},}\ }\bibinfo {note}
  {{http://tensorflow.org}}\BibitemShut {NoStop}%
\bibitem [{\citenamefont {Paszke}\ \emph {et~al.}(2019)\citenamefont {Paszke},
  \citenamefont {Gross}, \citenamefont {Massa}, \citenamefont {Lerer},
  \citenamefont {Bradbury}, \citenamefont {Chanan}, \citenamefont {Killeen},
  \citenamefont {Lin}, \citenamefont {Gimelshein}, \citenamefont {Antiga},
  \citenamefont {Desmaison}, \citenamefont {Kopf}, \citenamefont {Yang},
  \citenamefont {DeVito}, \citenamefont {Raison}, \citenamefont {Tejani},
  \citenamefont {Chilamkurthy}, \citenamefont {Steiner}, \citenamefont {Fang},
  \citenamefont {Bai},\ and\ \citenamefont {Chintala}}]{pytorch}%
  \BibitemOpen
  \bibfield  {author} {\bibinfo {author} {\bibfnamefont {Adam}\ \bibnamefont
  {Paszke}}, \bibinfo {author} {\bibfnamefont {Sam}\ \bibnamefont {Gross}},
  \bibinfo {author} {\bibfnamefont {Francisco}\ \bibnamefont {Massa}}, \bibinfo
  {author} {\bibfnamefont {Adam}\ \bibnamefont {Lerer}}, \bibinfo {author}
  {\bibfnamefont {James}\ \bibnamefont {Bradbury}}, \bibinfo {author}
  {\bibfnamefont {Gregory}\ \bibnamefont {Chanan}}, \bibinfo {author}
  {\bibfnamefont {Trevor}\ \bibnamefont {Killeen}}, \bibinfo {author}
  {\bibfnamefont {Zeming}\ \bibnamefont {Lin}}, \bibinfo {author}
  {\bibfnamefont {Natalia}\ \bibnamefont {Gimelshein}}, \bibinfo {author}
  {\bibfnamefont {Luca}\ \bibnamefont {Antiga}}, \bibinfo {author}
  {\bibfnamefont {Alban}\ \bibnamefont {Desmaison}}, \bibinfo {author}
  {\bibfnamefont {Andreas}\ \bibnamefont {Kopf}}, \bibinfo {author}
  {\bibfnamefont {Edward}\ \bibnamefont {Yang}}, \bibinfo {author}
  {\bibfnamefont {Zachary}\ \bibnamefont {DeVito}}, \bibinfo {author}
  {\bibfnamefont {Martin}\ \bibnamefont {Raison}}, \bibinfo {author}
  {\bibfnamefont {Alykhan}\ \bibnamefont {Tejani}}, \bibinfo {author}
  {\bibfnamefont {Sasank}\ \bibnamefont {Chilamkurthy}}, \bibinfo {author}
  {\bibfnamefont {Benoit}\ \bibnamefont {Steiner}}, \bibinfo {author}
  {\bibfnamefont {Lu}~\bibnamefont {Fang}}, \bibinfo {author} {\bibfnamefont
  {Junjie}\ \bibnamefont {Bai}}, \ and\ \bibinfo {author} {\bibfnamefont
  {Soumith}\ \bibnamefont {Chintala}},\ }\bibfield  {title} {\enquote {\bibinfo
  {title} {Pytorch: An imperative style, high-performance deep learning
  library},}\ }in\ \href
  {http://papers.neurips.cc/paper/9015-pytorch-an-imperative-style-high-performance-deep-learning-library.pdf}
  {\emph {\bibinfo {booktitle} {Advances in Neural Information Processing
  Systems 32}}},\ \bibinfo {editor} {edited by\ \bibinfo {editor}
  {\bibfnamefont {H.}~\bibnamefont {Wallach}}, \bibinfo {editor} {\bibfnamefont
  {H.}~\bibnamefont {Larochelle}}, \bibinfo {editor} {\bibfnamefont
  {A.}~\bibnamefont {Beygelzimer}}, \bibinfo {editor} {\bibfnamefont
  {F.}~\bibnamefont {d\textquotesingle Alch\'{e}-Buc}}, \bibinfo {editor}
  {\bibfnamefont {E.}~\bibnamefont {Fox}}, \ and\ \bibinfo {editor}
  {\bibfnamefont {R.}~\bibnamefont {Garnett}}}\ (\bibinfo  {publisher} {Curran
  Associates},\ \bibinfo {year} {2019})\ p.\ \bibinfo {pages}
  {8024}\BibitemShut {NoStop}%
\bibitem [{foo()}]{footnote}%
  \BibitemOpen
  \href@noop {} {}\bibinfo {howpublished} {The optimization problem of the
  network is linear, but it is not trivial to reduce it to a system of linear
  equations $\mathbf{A}\mathbf{x}=\mathbf{b}$; thus, we use the standard
  approach of training the network that is quick and efficient.}\BibitemShut
  {Stop}%
\bibitem [{\citenamefont {Szafran}\ \emph {et~al.}(1999)\citenamefont
  {Szafran}, \citenamefont {Adamowski},\ and\ \citenamefont
  {Bednarek}}]{SZAFRAN1999}%
  \BibitemOpen
  \bibfield  {author} {\bibinfo {author} {\bibfnamefont {Bart\l{}omiej}\
  \bibnamefont {Szafran}}, \bibinfo {author} {\bibfnamefont {Janusz}\
  \bibnamefont {Adamowski}}, \ and\ \bibinfo {author} {\bibfnamefont
  {Stanis\l{}aw}\ \bibnamefont {Bednarek}},\ }\bibfield  {title} {\enquote
  {\bibinfo {title} {Electron-electron correlation in quantum dots},}\ }\href
  {\doibase https://doi.org/10.1016/S1386-9477(99)00039-9} {\bibfield
  {journal} {\bibinfo  {journal} {Phys. E}\ }\textbf {\bibinfo {volume} {5}},\
  \bibinfo {pages} {185} (\bibinfo {year} {1999})}\BibitemShut {NoStop}%
\end{thebibliography}%

\end{document}